\documentclass[]{aa}

\usepackage{caption, subcaption}
\usepackage{txfonts}
\usepackage{amsmath}
\usepackage{epsfig}
\usepackage{graphicx}
\usepackage{color}
\usepackage[normalem]{ulem}
\usepackage{amsfonts}
\usepackage{xspace}
\usepackage{xcolor}
\usepackage[breaklinks, plainpages=false, colorlinks=true, anchorcolor=cyan, linkcolor=red, citecolor=cyan, urlcolor=magenta, bookmarks=false]{hyperref}

\newcommand{\params}{\mathbf\lambda} 
\newcommand{\Params}{\mathbf\Lambda}
\newcommand{\amp}{\mathcal{A}}
\newcommand{\f}{f_0}
\newcommand{\phase}{\varphi_0}
\newcommand{\fdot}{\dot{f}}

\newcommand{\data}{{\bf d}}

\newcommand{\Nparams}{N_{\rm P}}
\newcommand{\Nsources}{N_{\rm GW}}
\newcommand{\template}{{\boldsymbol h}}
\newcommand{\templatesum}{{\mathbf h}}
\newcommand{\innerproduct}[2]{\langle #1 | #2 \rangle}

\begin{document}
\renewcommand{\thefigure}{\arabic{figure}}
\setcounter{figure}{0}

 \def\I{{\rm i}}
 \def\E{{\rm e}}
 \def\D{{\rm d}}

\title{The LISA Data Challenge $Radler$ Analysis} 
\subtitle{and Time-dependent Ultra-compact Binary Catalogues}

   \author{Kristen Lackeos
          \inst{1}\fnmsep\inst{2}\fnmsep\thanks{klackeos@mpifr-bonn.mpg.de},
          Tyson B. Littenberg\inst{2}, 
          Neil J. Cornish\inst{3},
          James I. Thorpe\inst{4}
          }

   \institute{Max-Planck-Institut f{\"u}r Radioastronomie (MPIfR), Auf dem H{\"u}gel 69, 53121, Bonn, Germany
        \and
             NASA Marshall Space Flight Center, Huntsville, Alabama 35811, USA
        \and
             eXtreme Gravity Institute, Department of Physics, Montana State University, Bozeman, Montana 59717, USA
        \and
             NASA Goddard Space Flight Center, Greenbelt, Maryland 20771, USA
             }

   \date{Received June 18, 2023; accepted }




  \abstract
    {Galactic binaries account for the loudest combined continuous gravitational wave signal in the Laser Interferometer Space Antenna (LISA) band, which spans a frequency range of 0.1 mHz to 1 Hz.}
    {A superposition of low frequency Galactic and extragalactic signals and instrument noise comprise the LISA data stream. Resolving as many Galactic binary signals as possible and characterising the unresolved Galactic foreground noise after their subtraction from the data are a necessary step towards a global fit solution to the LISA data.}
    {We analyse a simulated gravitational wave time series of tens of millions of ultra-compact Galactic binaries hundreds of thousands of years from merger. This data set is called the $Radler$ Galaxy and is part of the LISA Data challenges.  We use a Markov Chain Monte Carlo search pipeline specifically designed to perform a global fit to the Galactic binaries and detector noise. Our analysis is performed for increasingly larger observation times of 1.5, 3, 6 and 12 months.}
    {We show that after one year of observing, as many as ten thousand ultra-compact binary signals are individually resolvable. Ultra-compact binary catalogues corresponding to each observation time are presented. The $Radler$ Galaxy is a training data set, with binary parameters for every signal in the data stream included. We compare our derived catalogues to the LISA Data challenge $Radler$ catalogue to quantify the detection efficiency of the search pipeline.  Included in the appendix is a more detailed analysis of two corner cases that provide insight into future improvements to our search pipeline.} 
    {}


\maketitle

\section{\label{sec:intro}Introduction}
Ultra-compact binaries (UCBs) are compact or degenerate star systems with orbital periods of a few hours or less. They emit continuous gravitational radiation with frequencies in the mHz range. Circularised compact binaries of the Milky Way Galaxy are expected to be the most numerous type of gravitational wave (GW) signal below $\sim 5$ mHz. Double white dwarfs (WDs) are the most common type of UCB, although UCBs can also involve neutron stars or black holes, some possibly with non-zero eccentricity.

Here we analyse a simulated data set from a future space-based GW detector LISA~\citep{hi90,LISA} and present time-evolving UCB catalogues for 1.5, 3, 6, and 12 months of simulated data. We resolve as many as 10,000 binaries from a 12-month observing period. More than 400 of these are constrained to a sky localisation area of 10 deg$^2$ or better. The simulated data span a 24-month period. Here, we do a full analysis up to 12 months. In the Appendix we discuss a few cases which have poor convergence of the sampler after 12 and 24 month analyses. These corner cases were identified when comparing the the 6 and 12-month catalogues. 

LISA is a European Space Agency led mission, in collaboration with NASA and an international consortium of scientists~\footnote{https://lisa.pages.in2p3.fr/consortium-userguide/}, designed to explore the uncharted Universe of low-frequency GWs and promises `answers to fundamental questions in physics and astronomy.'\footnote{Astro2020, Pathways to Discovery in Astronomy and Astrophysics for the 2020s, National Academies of Sciences, Engineering and Medicine, Washington, DC: The National Academies Press, 2021, https://nap.nationalacademies.org/resource/26141/interactive/}. Ground-based detectors are insensitive to low frequency GWs due to gravity gradient noise from terrestrial sources. One must use space-based detectors to observe frequencies below 1 Hz.

LISA will monitor the observable Universe with a triangular constellation of spacecraft separated by 2.5 million km. Each spacecraft houses two free flying test masses (TM) and two lasers linking the two other spacecraft. Heterodyne laser interferometry will be used to observe picometer level changes in TM separations~(\cite{Weise:2017}). The observatory lies in a plane inclined 60$^{\circ}$ with respect to the ecliptic and will be in a heliocentric orbit with a period of one year. This arrangement is sensitive to GWs spanning four decades in frequency from 0.1 mHz to 1 Hz~(\cite{baker19}). A sampling of LISA's most anticipated sources are a stochastic GW background (SGWB) from cosmological sources~(\cite{Christensen}), late-time coalescence and merger of massive black hole binaries (MBHB)~(\cite{Klein16}), and stellar (\cite{Sesana_2016}), intermediate and extreme mass ratio inspirals~(\cite{am07}). Finally, there will be the unanticipated or, yet, unknown astrophysical signals~(\cite{corn19}). LISA will observe this multitude of overlapping signals simultaneously. 

Resolving UCB signals from a noisy LISA data stream is the focus of this paper. Tens of millions of UCBs are expected to emit GWs below $\sim$ 5 mHz, with the overwhelming majority forming a Gaussian confusion, or foreground, noise in excess of LISA's instrumental noise. At higher frequencies, where there are fewer UCBs, the foreground is reduced substantially \citep{ni12,Nelemans01}. For frequencies below $\sim$ 5 mHz, however, characterisation of the UCB foreground component will be essential to disentangling the SGWB, transient, and continuous extragalactic signals that overlap in frequency and time with each other and with the numerous UCB signals. It is necessary to perform a global fit \citep{Crowder:2006eu,CornCrow15} to the data, where the GW signals and noise sources are fit simultaneously and the number of GW signals in the data is an unknown variable. A LISA global fit pipeline for UCB parameter estimation has been in development for several years (\cite{lit20}) and has recently been extended to include MBHBs (\cite{Littenberg:2023xpl}). 

The generation and analysis of simulated LISA data streams is driven by a series of data challenges issued to the broader data analysis community. The first collection of pipelines were developed as part of the Mock LISA Data Challenge \citep[MLDC;][]{Arnaud07,ArnBab,Babak_2008,BabBak,BabBak2,Babak_2010}. The MLDC was designed to facilitate and coordinate the development of data analysis pipelines, and signal waveforms. The most recent incarnation of this effort, begun in the Spring of 2018, is now known as the LDC\footnote{https://lisa-ldc.lal.in2p3.fr/ldc}, with an aim to improve existing algorithms, and create new ones, to generate a common platform to evaluate and compare the performance of different algorithms, to address science requirements, and with overarching goal of developing `mission ready' end-to-end data analysis pipelines. 

The first LDC data set $Radler$ comprises four separate challenges, each focusing on extracting a different type of gravitational wave source from a noisy data stream: stochastic signals, single MBHB, single EMRI, and a Galactic binary (GB) or UCB population. The latter is referred to as `the Galaxy' \footnote{https://lisa-ldc.lal.in2p3.fr/challenge1}. The LDC Galaxy was constructed from the simulations of \citet{Toonen:2017yct}. In this paper we report on the analysis of the LDC Radler Galaxy using data set LDC1-4$\_$GB$\_$v1. 

In recent years, a number of techniques have been developed for the analysis of gravitational waves from a simulated population of UCBs. Using the $Radler$ verification binary data set as a starting point \citet{Strub:2022upl} inject overlapping signals and use differential evolution to find a maximum likelihood estimate (MLE) and then apply the Metropolis-Hastings algorithm to sample the posterior distribution around the MLE of each candidate detection to determine parameter uncertainties. To generate posteriors, \citet{Strub:2022upl} uses Gaussian Process Regression with the existing `FastGB' LISA response simulation (\cite{Cornish:2007if}) to model the log-likelihood function. This helps further decrease computation time, since the approximated LISA response is not simulated for each sample. We say more about `FastGB' in Section \ref{sec:code}, when we introduce our signal model. The authors of \citet{ZhangXH:2021} also use MLE, but with particle swarm optimisation and evaluation of the $\mathcal{F-}$statistic to resolve UCBs. Different techniques have their own strengths and weaknesses. For multi-modal distributions, MLE is vulnerable to missing the modes which contain the true parameter values. However, for targeted searches of known EM binaries (for example, known in optical, X-ray, $\gamma$-ray, or radio) this limitation can be mitigated by designing priors informed by the EM observations.

Thousands of detached-binaries and a few ten to hundreds of interacting systems will be detected (\cite{ni12}). These systems provide unique laboratories for fundamental physics. For example new bounds on the graviton mass competitive with those produced by ground based GW observatories and pulsar timing arrays are expected using eclipsing (\cite{co04}) and eccentric (\cite{jo05}) compact binaries. Modifications to GR in low velocity regimes can be constrained with high mass white dwarf binaries (\cite{Littenberg_2019}). With an observed population of relativistic binaries the Milky Way (MW) potential will be mapped using GWs (\cite{PhysRevD.86.124032}), and regions previously obscured by intervening material will be revealed (\cite{Korol2018}). Models of MW globular cluster formation and evolution will be further constrained (\cite{be13}).  In \citet{da19} the authors discuss the prospect of gravitational waves affirming the existence of post-main sequence exoplanet and brown dwarf populations in the MW. \cite{br20} demonstrate that a catalogue of thousands of MW UCBs will help constrain binary star formation and evolution. Finding a WDs binary system near the threshold of merging would reveal new insights into the precursor physics of Type Ia supernova (\cite{webb10}). Multi-messenger astrophysics of compact binary stars will be enriched with a GW perspective. Follow-up electromagnetic (EM) searches of newly detected LISA binaries will confirm relativistic binaries missed by traditional MW globular cluster searches \citep{kr18,kr19}. For known multi-messenger binaries, that is systems that have been observed electromagnetically, joint EM-GW observations will provide improved physical constraints on masses, orbital parameters and dynamics, beyond what independent EM or GW observations achieve on their own \citep{sh14,li19b}.  On classes of UCBs unobserved electromagnetically, \citet{sb21} predict with simulation and semi-analytic evolution models that WD-black hole binaries will be detectable and could inform follow-up searches in X-ray. LISA detections of Galactic black hole/white dwarf-neutron star binaries will inform and increase the computational efficiency of radio searches for pulsars in these systems \citep{ky19,th20}.
It is also possible to use UCBs as phase/amplitude standards for self-calibration of the data~(\cite{Littenberg_2018}). There is also a technique to use the WD binary annual modulation to extract an isotropic astrophysical SGWB (\cite{PhysRevD.89.022001}, \cite{LinS:2022}), which depends on first resolving and remove as many UCBs as possible. There is even a proposal to use UCBs as a GW timing array to indirectly detect GWs in the low frequency regime (nHz to $\mu$Hz) (\cite{br22})! Looking even further to the future, UCB analysis has been investigated using a coherent network of at least three independent space-based gravitational wave detectors (\cite{ZhangXH:2022}).

Gravitational waves from individual and populations of UCBs are interesting in their own right, $\emph{i.e.}$ beyond considering them a foreground noise source. Their characterisation and extraction is an integral part of the `global fit' solution (\cite{lit20}) for extragalactic source detection at mHz frequencies. Listed above are just a few reasons why the analysis presented here is vital to achieving the widest possible scientific impact for the LISA mission. The main motivation of this paper is to further test and develop the Galactic Binary Markov Chain Monte Carlo (GBMCMC) search pipeline in preparation for a global fit.

The individual sections of our paper are as follows. In Section~\ref{sec:code} we provide an introduction to the likelihood used in our analysis, the noise and signal models, and the UCB parameterisation. The computational resources used are also discussed there. We present GBMCMC catalogues for the observation times analysed in Section~\ref{sec:catalogue}. For each catalogue we make various signal-to-noise ($S/N$) cuts to the data, and identify well-localised UCBs to target for multi-messenger studies. In addition to resolving as many UCBs as possible, we quantify the efficacy of our search pipeline by comparing the GBMCMC catalogue to the LDC $Radler$ UCB catalogue, classifying each catalogue UCB as either matched, confused or false alarm (Section~\ref{sec:ldc_inj}). Lastly, we summarise our results in Section~\ref{sec:discussion}. In Appendix \ref{sec:app}, two catalogue UCBs classified as confused are examined in more detail. These serve as case studies for future development of the sampler.  

\section{\label{sec:code}GBMCMC analysis}

The GBMCMC search pipeline (\cite{lit20}) uses Bayesian model selection to optimise the number of detectable UCBs. In a nutshell, GBMCMC performs a global fit to the resolvable binaries using a trans-dimensional (reversible jump)~(\cite{doi:10.1093/biomet/82.4.711}) MCMC algorithm with parallel tempering~(\cite{PhysRevLett.57.2607}). At the same time, it fits a model to the residual confusion noise. 

Parallel tempering is used to prevent the sampler from becoming trapped in sub-dominate modes of the posterior, by sampling with parallel chains of different temperatures, with exchanges of parameters between chains subject to detailed balance. Higher temperature chains are more freely able to sample the parameter space. For example, a chain given an infinite temperature will simply sample the prior distribution~(\cite{Littenberg:2010}). 

Trans-dimensional MCMC algorithms addresses the model selection aspect of the problem. The MCMC stochastically transitions between models, where each model contains a different number of UCBs, while satisfying detailed balance. Therefore, the number of iterations the chain spends in a particular model is proportional to the marginalised likelihood, or evidence, for that model. Before saying more about the waveform and noise models, we describe a likelihood for Bayesian inference adapted for LISA science analysis. 

For our analysis we used 100,000 MCMC steps, after the burn-in phase. Convergence time depends critically on sampling from customised proposal distributions (\cite{lit20}). In the Appendix we discuss a few cases which have poor convergence of the sampler after 12 and 24 month analyses.

\subsection{Likelihood and noise constructions}

The three LISA spacecraft communicate with each other via laser links forming the interferometric arms of the detector.
The arms of a space-based detector will have different lengths, varying on the order of a few percent of their length over the course of a year. This occurs due to the solar wind, the gravitational coupling of the Earth-moon system and the influence of the other planets in the solar system on the spacecraft orbits, causing the test masses to deviate from their Keplerian orbits. In an equal arm detector, laser frequency noise experiences the same delay in each arm and will cancel at the detector. For time-varying arm-lengths, Time-Delay Interferometry \citep{Prince:2002hp, Adams:2010vc} (TDI) has been developed to algorithmically remove the otherwise dominating laser frequency noise by generating virtual equal-armlength interferometers, performed on the ground in post-processing. In general, many different TDI combinations of interferometer output signals, or observables, are possible (\cite{est00}). For the LISA mission two quasi-independent Michelson interferometer data streams and a third null-stream (the LISA “Sagnac” observable) will be constructed in post-processing\footnote{https://www.cosmos.esa.int/documents/678316/1700384/SciRD.pdf}. 

The likelihood function \eqref{likelihood} depends on the TDI observables, or `channels', used in the algorithm. The LISA signal is a superposition of two parts: the frequency response of the $I^{\rm th}$ channel to all the gravitational wave signals incident on the detector, ${\bf h}_I$, and the combination of all the noise sources impacting that channel, ${\bf n}_I$: $\data_I = {\bf h}_I + {\bf n}_I$. The ``noise'' term is a superposition of instrument noise and gravitational wave signals that are individually too quiet to extract from the data (forming a confusion noise below $\sim$ 5 mHz). The detectable gravitational wave signal is recovered using a signal model ${\bf h}_I$ such that the residual ${\bf r}_I = \data_I - {\bf h}_I$ is consistent with the noise model. For Gaussian noise the likelihood is:
\begin{equation}
p(\data | {\bf h}) = \frac{1}{(2\pi \, \det{\bf C})^{1/2}} \, \exp\left(- \frac{1}{2}(d_{Ik} - h_{Ik}) C^{-1}_{(I k)(J m)} (d_{Jm} - h_{Jm})\right)\ \label{likelihood},
\end{equation}
where ${\bf C}$ is the noise correlation matrix. Indices $k$ and $m$ correspond to the data samples for the $I$ and $J$ channels respectively, where there is an implicit sum over the TDI channels $I=\{X,Y,Z\}$ and data samples. 

If the fluctuations in the data are stochastic the noise correlation matrix in terms of frequency becomes partially diagonalised $C_{(I k)(J m)} = S_{IJ}(f_k) \delta_{km}$, where $S_{IJ}(f)$ is the cross-power spectral density between channels $I,J$~\citet{Adams:2010vc}. Since the noise levels are equal and uncorrelated on each spacecraft, noise orthogonal TDI variables $I'=\{A,E,T\}$~(\cite{Prince:2002hp}) are constructed such that the cross-spectral density matrix is diagonalised by performing a linear transformation in the space of TDI variables. See \citet{Adams:2010vc} for complete expressions for the instrument noise contributions to the cross spectra $S_{IJ}(f)$, where the realistic scenario of unequal noise levels in each spacecraft is treated.

The $\{A,E,T\}$ combination also results in signal orthogonality for frequencies below the inverse round-trip light travel time along the arms of the instrument (c/2$\pi$L), $f_*\simeq 19.1 \; {\rm mHz}$, such that $A \sim h_+$, $E \sim h_\times$ (corresponding to the two virtual Michelson interferometer channels) and $T \sim h_\odot$ (the null channel). This is not restricted to equal arms; \citet{Adams:2010vc} derived a combination that maintains this insensitivity for unequal arm length detectors. The gravitational wave response of the null $T$ channel is highly suppressed for $f < f_*$. For this reason the Sagnac data combination is particularly valuable for noise characterisation and the detection of stochastic backgrounds \citep{Tinto:2001ii,Hogan:2001jn} and unmodelled signals~(\cite{travis4}).  

For this analysis we have made a number of simplifying assumptions, to be relaxed in the future. We take the noise to be stationary and assume the noise correlation matrix is diagonal in the frequency domain. In reality, the confusion noise is cyclo-stationary, with periodic amplitude modulations imparted by LISA's orbital motion~(\cite{PhysRevD.69.123005}). Here we neglect off-diagonal terms in the frequency domain noise correlation matrix ${\bf C}$. Since an overwhelming majority of signals have frequencies well below frequency $f_*$, we only use the $A$ and $E$ data combinations in the analysis and we assume that the noise in these channels is uncorrelated. The instrument response includes finite arm-length effects of the LISA constellation and arbitrary spacecraft orbits. The TDI prescription currently implemented treats the arm lengths as equal and unchanging with time, saving on computational cost. We split the analysis into 3317 sub-bands $[f_i, f_i+N/T_{\rm obs}]$, where $f_{i,\rm{min}}$ = 0.00813 mHz. The noise in each band is approximate as an undetermined constant $S_i$. The noise level in each band becomes a free parameter explored by the Reversible-jump Markov chain Monte Carlo algorithm, resulting in a piece-wise fit to the instrument noise over the full analysis band. 

\subsection{Compact binary parameterisation and signal model}
A compact binary orbit is modelled with eight parameters, $\Nparams=8$, $\params \rightarrow (\amp, \f, \fdot, \phase, \iota, \psi, \theta, \phi)$, where $\amp$ is the amplitude, $\f$ is the observed GW frequency, which is twice the orbital frequency of the binary, $\fdot$ is the (constant) time derivative of the GW frequency, $\phase$ is the initial GW phase at the observation start time, $\iota$ is the inclination of the orbital plane relative to the line of sight, the wave polarisation axes in the Solar System barycentre are determined by $\psi$, and $\theta,\phi$ are the ecliptic latitude and longitude, respectively. See \citet{Shah2012} for a description of parameter correlations and degerneracies.

When at least 90 $\%$ of $\fdot$ MCMC samples are positive, we take the orbital evolution to be GW-dominated evolution and use \eqref{fdot_amp} to estimate the chirp mass ${\cal M}$ and luminosity distance ($D_L$) of the binary.

\begin{eqnarray}\label{MD}
\fdot &=& \frac{96}{5} \pi^{8/3} {\cal M}^{5/3} \f^{11/3}  \nonumber \\
\amp &=& \frac{2 {\cal M}^{5/3} (\pi \f)^{2/3}}{D_L} \label{fdot_amp}
\end{eqnarray}

We also have optional settings to include the second derivative of the frequency~(\cite{Littenberg_2019b}) as an additional parameter, in which case the frequency derivative is no longer constant, so the parameter $\fdot\rightarrow\fdot_0$ is fixed at the same fiducial time as $\f$ and $\phase$. 

The detector response of the $I^{\rm th}$ data channel to the signal from a galactic binary with parameters $\params_a$ is $\template_I(\params_a)$, and the superposition of individual UCBs forms the signal model:
\begin{equation}
\templatesum_I(\Params) = \sum_{a=1}^{\Nsources} \template_I(\params_a).
\end{equation}
The number of binaries in each sub-band, $\Nsources$, is {\it a priori} unknown, and has to be determined from the analysis. A probability distribution for $\Nsources$ is established, essentially producing a catalogue for each dimension. Here we build our catalogue with the dimension having the highest Bayesian evidence.
Individual binary systems are modelled as isolated point masses on slowly evolving quasi-circular orbits. Orbital eccentricity~(\cite{Seto:2001pg}), tides~(\cite{2012MNRAS.421..426F}) and third bodies~(\cite{Robson:2018svj}) are not included in our model. The signals are modelled using leading order post-Newtonian waveforms, and the instrument response ${\bf h}_I$ is computed with the fast-slow decomposition method, in terms of frequency~(\cite{Cornish:2007if}). FastGB, or the ``fast-slow'' method, is the decomposition of the relative path length variation of the detector arm, $\delta l(t)/L$. Namely, the product between the rapidly varying $\exp({i\omega_0t})$, where $\omega_0$ is the instantaneous angular frequency of the GW, and a slowly varying amplitude factor that depends on the LISA spacecraft orbits and the GW amplitudes, that is part of the LISA instrument response. 

The LISA instrument response is modulated as the detector rotates about its centre and orbits the Sun. The sensitivity pattern is anisotropic, so as the detector moves the sensitivity pattern evolves with time with respect to a given source. This modulation is imprinted on the detected amplitude. Additionally, as LISA orbits around the Sun, the frequency of the GW is Doppler-shifted, resulting in a time-dependent phase shift of the instrument response (\cite{Peterseim:1997ic}). Both effects introduce a spread in the power of the source such that it is no longer monochromatic when viewed from a LISA-based frame and depends on the direction and orientation of the source. The power is reduced relative to the instrumental noise as it is spread over a series of side-bands, offset from the GW frequency at integer multiples of the modulation frequency $f_m=(1\ \rm{yr})^{-1}.$ The subdominant harmonics lead to secondary maxima in the likelihood surface which are dealt with using tailored proposals \citep{Crowder:2006eu,Littenberg:2011zg}.
The detector velocity with respect to the Solar System barycentre evolves with time. Therefore, the phase modulation depends on a coupling between frequency and sky location. This effect helps one localise the source on the sky and determine its orientation, because each source has a unique modulation pattern. GBMCMC implements multi-modal proposals to account for degeneracies and symmetries in the likelihood surface to improve chain convergence time (\cite{lit20}). 

\subsection{Data segmentation and computational resources}
\begin{figure*}[h]
\begin{center}
\includegraphics[width=7.5in]{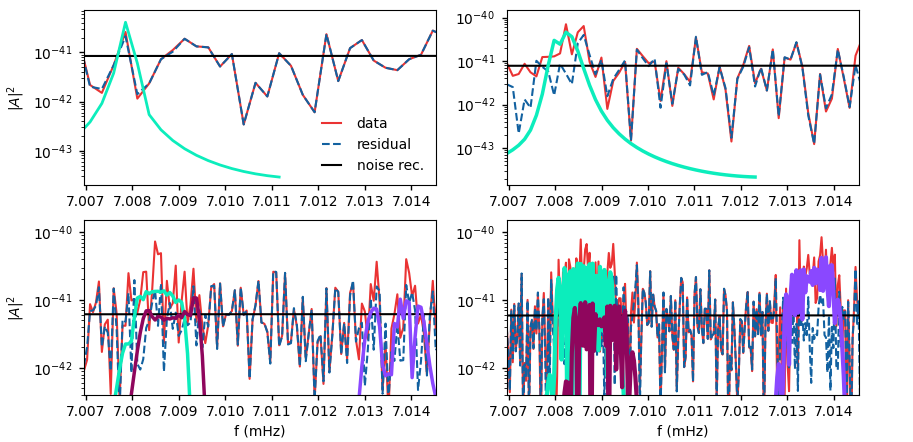}
\end{center}
\vspace*{-2mm}
\caption{Waveform evolution as a function of observing time. Individual waveforms are plotted over the input A-channel PSD (solid, red curve), noise (constant, black curve stretching across the frequency window) and residual (dashed, blue curve) curves. The top graphs show a single 1.5-month (left panel) and 3-month (right panel) catalogue waveform, in green. The bottom panels show, in total, three 6 and 12-month catalogue waveforms, two of which did not appear in any earlier catalogue. All injected signals in this frequency window were recovered in a 6 and 12 month analysis.}
\label{waveform_time_ev}
\end{figure*}

Four time periods were searched over: $T_{\rm obs}=$ 1.5, 3, 6 and 12 months, each with the same starting time. A catalogue is produced for each time period searched. The catalogue data include a point estimate of the UCB parameters, the waveforms, and posterior distributions for the parameters $\lambda$. The search is done in terms of frequency, where the full LISA band is divided into 3317 `analysis segments' or `frequency segments'. The LISA band is divided into frequency segments of width $2^5$/$T_{\rm 1.5mo}$, $2^6$/$T_{\rm 3mo}$, $2^7$/$T_{\rm 6mo}$ and $2^8$/$T_{\rm 12mo}$ for the 1.5, 3, 6 and 12-month analyses, respectively. The $f_{\rm min}$ and $f_{\rm max}$ for each of the 3317 analysis segments are the same for each $T_{\rm obs}$. In Figure \ref{waveform_time_ev}, we see how the waveforms in a particular frequency segment evolve with observing time. After 12 months of analysis all of the LDC $Radler$ UCB signals in this segment have been recovered. An example of frequency spreading due to detector motion is also apparent in Figure \ref{waveform_time_ev}, where the waveforms do not appear monochromatic. 

Each analysis segment is padded with data amounting to the typical bandwidth of a source. This creates a certain amount of overlap with neighbouring analysis segments. This allows the MCMC to explore the
data in the padded region, which is especially useful for sources with long tails that extend beyond the hard boundary of the analysis segment. During catalogue production, only samples fitting sources in the original analysis window are retained. This prevents the same source from appearing in the catalogue more than once.
In the next section, we discuss how the raw chain samples from the MCMC analysis are sorted into individual UCB catalogue entries and present the results of our analysis in the form of evolving catalogues as a function of $T_{\rm obs}.$

To perform the wholesale analysis of $\sim$ 3000 frequency segments in a reasonable amount of time, we used Amazon Web Service (AWS) cluster computing resources \footnote{http://aws.amazon.com/what-is-aws/}. Each segment was analysed in an `embarrassingly parallel' way, such that there is no communication between segments being analysed. There is interest in using cloud-based computing resources for the actual LISA mission, so our analysis is a first test run in using this infrastructure. 

GBMCMC and the software used to produce the catalogues are downloadable from GitHub~(\cite{ldasoft}). Our catalogue data are available upon request. Additionally, a Python package dedicated to exploring these data is available~(\cite{lisacat}).

\section{\label{sec:catalogue}The recovered GBMCMC-Radler catalogue}

The development of UCB catalogues for the LISA mission is the process of transforming the search data products into a form that is useful to the greater astronomy community. In our case, this means filtering the GBMCMC parameter chain outputs into individual catalogue entries using the maximum likelihood model. Before moving on to the catalogue results, we summarise the process of filtering posterior samples to construct catalogue entries. The details of this process are also found in \citet{lit20} and \citet{Littenberg:2023xpl}. 

To build a catalogue for a particular frequency segment, we start with the highest evidence model chain. Namely we select the $\Nsources$-source model which has the highest evidence. The correlation, or overlap, value between waveforms $\template(\params_i)$ and $\template(\params_j)$ for binaries with parameters, $\params_i$ and $\params_j$, Equation (\ref{eq:match}), is used as a metric to cluster parameter samples.
\begin{equation}\label{eq:match}
M_{ij} \equiv \frac{\innerproduct{\template(\params_i)}{\template(\params_j)}}{\sqrt{ \innerproduct{\template(\params_i)}{\template(\params_i)} \innerproduct{\template(\params_j)}{\template(\params_j)} }}.
\end{equation}
We set a threshold $M_*=0.5$ above which parameter sets are interpreted as describing the same UCB template. 

The first $\Nsources$ samples in the chain correspond to the first $\Nsources$ UCB catalogue entries, in no particular order. Waveforms for these first $\Nsources$ entries are generated and used as reference entries for cross-correlating with other samples waveforms. If the correlation value between a given sample waveform and a reference entry waveform exceeds the default threshold value of $M_*$, the sample parameters are appended to the entry parameters. Correlations are only computed when a sample has an $\f$ that is within a range of $10/T_{\rm obs}$ bins of the reference entry $\f$. If a chain sample is not within range of an existing entry or does not have a correlation $M_{ij}>M_*$ with an existing entry, a new entry is created and added to the list of reference entries to be matched against. This process continues until all chain samples are grouped. 

 Each entry has an associated evidence that is used to further filter the number of entries. The evidence for an entry $p(\data) = \int p(\data|\params)\, d\params$ is proportional to the total number of chain samples. The evidence for an entry is computed as the number of chain samples in the entry divided by ($N_{\rm total}$/$\Nsources$), where $N_{\rm total}$ is the total number of samples in the chain. A threshold evidence of 0.5 must be exceeded for a particular entry to be included in the final catalogue.

The filtered parameter chains for each entry are used to form additional catalogue products. An entry's point-estimate is chosen to be the sample that corresponds to the median of the marginalised posterior on $\f$. We also compute the full multi-modal $\Nparams\times\Nparams$ covariance matrices for each mode of the posterior. These are then used for covariance matrix proposals as more data are acquired. From the point-estimates of each entry, waveform entries are computed. Finally, metadata about the catalogue are stored including the total number of above-threshold entries, their weights and S/N and the full set of posterior samples for each entry. History data are also included, which simply links catalogue entries with preceding $T_{\rm obs}$ catalogue entries, if such a link exists. This is dependent on the correlation of entry waveforms with preceding $T_{\rm obs}$ catalogue entry waveforms. A default threshold of $M_*=0.5$ is again used to make an association, but the user can adjust this as needed.

\begin{figure*}[h]
\begin{center}
\includegraphics[width=6.5in]{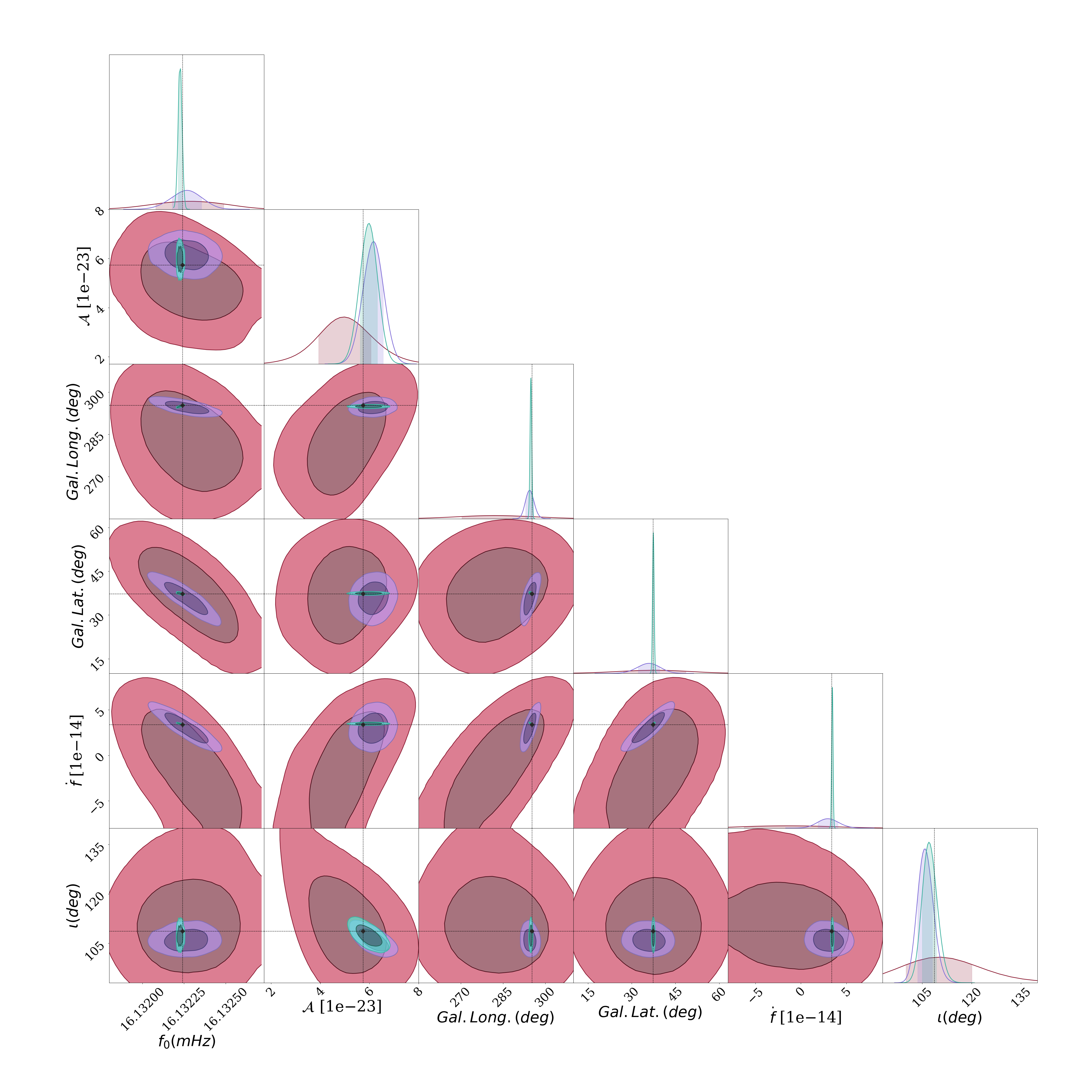}
\end{center}
\vspace*{-5mm}
\caption{A well-localised, eclipsing and chirping UCB for EM follow-up. The LDC parameter values are shown as black markers on the GBMCMC posteriors. The 1$\sigma$ and 2$\sigma$ posterior curves are graphed in this corner plot. Colours pink, purple, and green are the 3, 6, and 12 month posteriors, respectively. }
\label{sky1}
\end{figure*}

\begin{figure}[h]
\begin{center}
\includegraphics[width=3.5in]{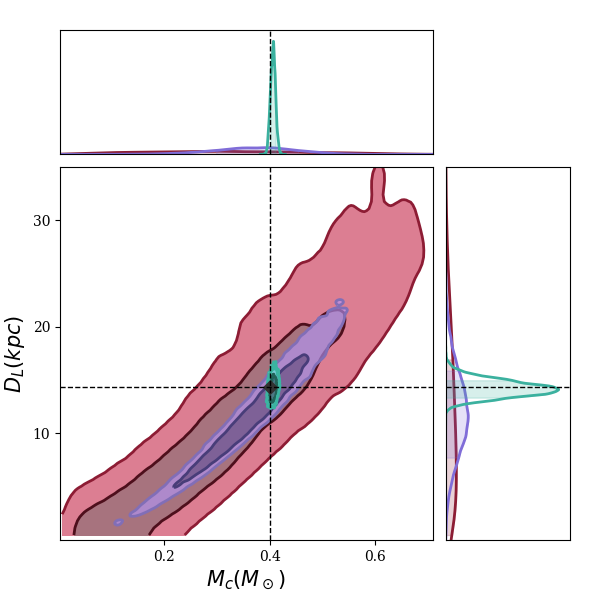}
\end{center}
\vspace*{-5mm}
\caption{Luminosity distance and chirp mass for a well-localised, eclipsing and chirping UCB. The 12-month posteriors $\dot f$ and $A$ the binary from Figure \ref{sky1} were re-sampled using Equation (\eqref{fdot_amp}) to form the luminosity distance $D_L$ and chirp mass $M_{\odot}$ posteriors. The black marker on the graph corresponds to the $D_L$ and $M_c$ derived by substituting the injected LDC parameters into \eqref{fdot_amp}. See Figure \ref{sky1} for a description of the three colours.}
\label{sky2}
\end{figure}

The numbers of catalogue UCBs for our 1.5, 3, 6 and 12 month LDC analyses are shown in Table \ref{table1}, along with UCB number as a function of different $S/N$ cuts in Table \ref{table2}. In Table \ref{table3} we make various parameter cuts on catalogue UCBs that have frequencies above 5 mHz. Catalogue UCBs with at least 90 $\%$ of their samples meeting the particular parameter cut criteria are included in the count. 
In Figure \ref{sky1} we show parameter posteriors, $\f$, $\amp$, sky location, $\fdot$, $\iota$, for a well-localised eclipsing UCB as a function of observing time. 
In Figure \ref{sky2}, we graph the chirp mass and luminosity distance posteriors using Equation (\eqref{fdot_amp}). The LDC values of chirp mass and distance, 0.4004$M_{\odot}$ and 14.39 kpc, are within 1$\sigma$ uncertainty of the the derived chirp mass and distance, $0.4073^{+0.0040}_{-0.0057} M_{\odot}$ and $14.12^{+0.86}_{-0.77}$kpc. 

\begin{figure*}[h]
\includegraphics[width=1.0\textwidth]{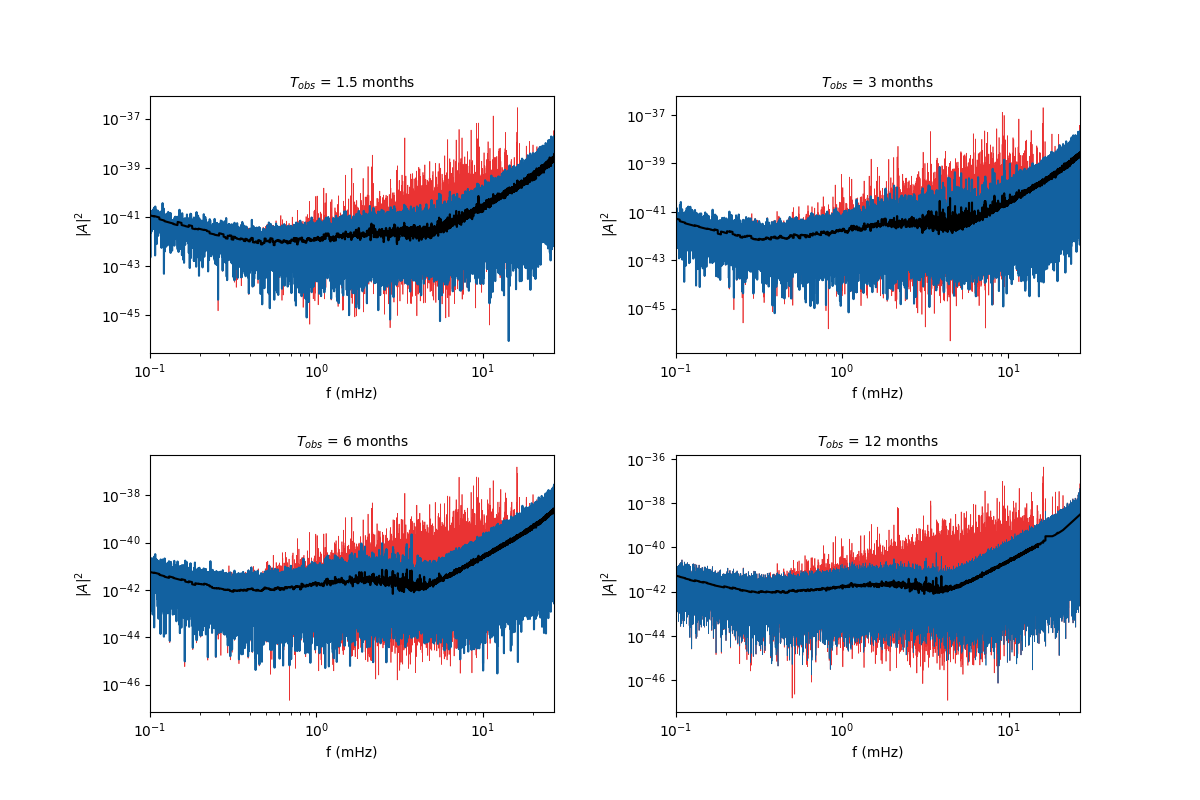} 
\caption{\label{fig:money_plot} The power spectra graphs for the analysed LISA band, for the 1.5, 3, 6, and 12-month catalogue. The red curve is the A-channel input data, and the dashed, blue curve is the residual, after the catalogue UCBs have been subtracted. The black curve plotted on top of the data and residual is the noise level.}
\label{figure2a}
\end{figure*}

\begin{figure*}[h]
\includegraphics[width=1.0\textwidth]{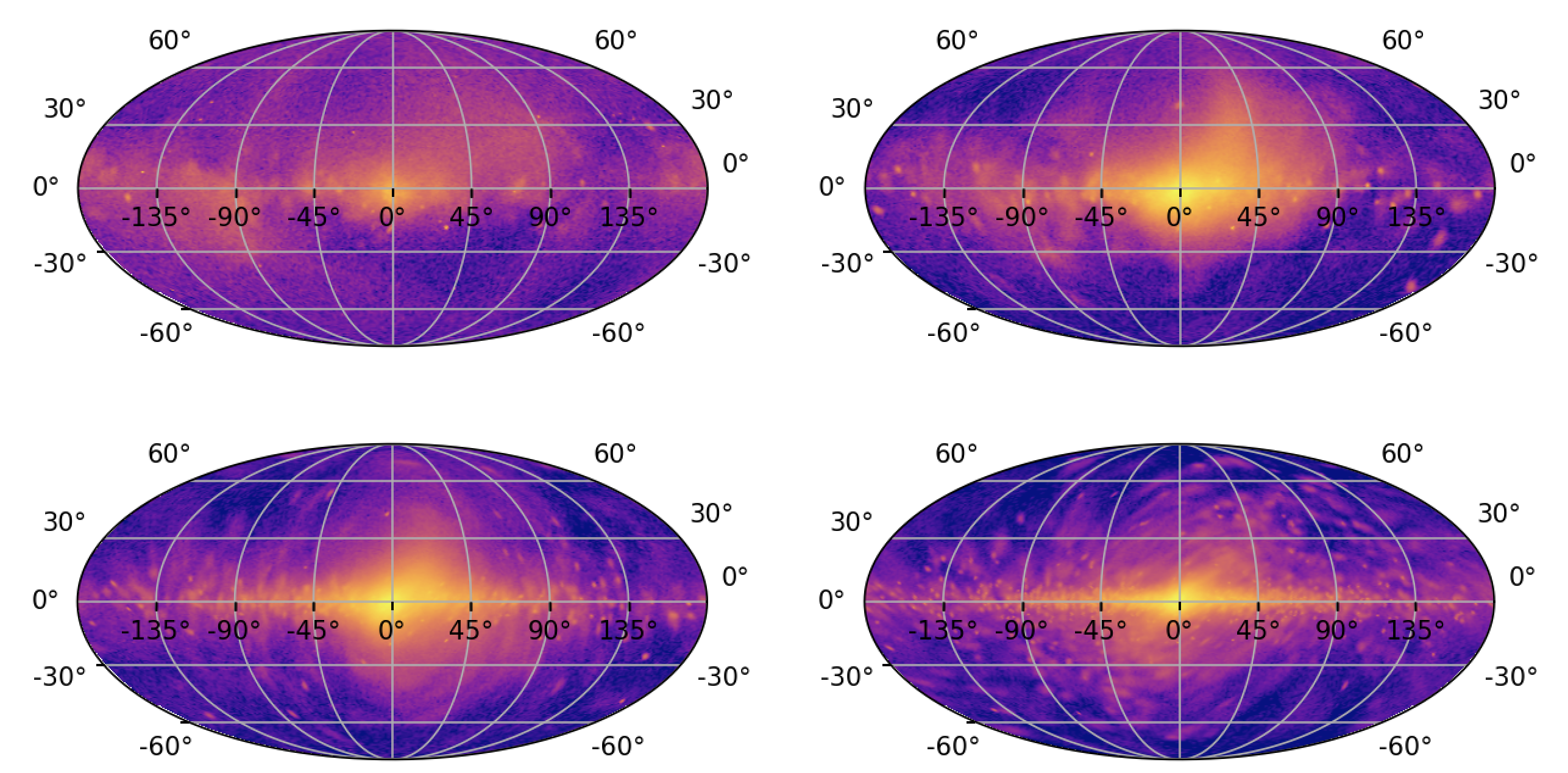} 
\caption{\label{fig:money_plot2} 
A Molweide projection of the sky location posteriors for 1.5, 3, 6, and 12-month catalogue UCBs in galactic coordinates, with the same ordering as Figure \ref{figure2a}. Every tenth posterior sample has been used to construct the sky location graphs.}
\label{figure2b}
\end{figure*}

In Figure \ref{figure2a} we show the power spectral density for the full frequency band that was analysed. The residual becomes smoother with time, and one can clearly see the excess confusion noise below $\sim$4 mHz. In Figure \ref{figure2b} are joint posteriors for sky location in ecliptic coordinates, for each observation time, 1.5, 3, 6, and 12-months, from left to right starting from the top. The sky location posteriors are graphed using every tenth sample.

One use of the catalogue data are posterior-based proposals for individual UCBs for use in future global fits to the LISA data. Updates to UCB parameters are proposed, independently of other UCBs in the catalogue. The cadence of applying new proposals to the global fit is to be determined for the LISA mission. 

For this analysis, we applied covariance matrix proposal updates to the 3, 6, and 12 month analysis, respectively using the 1.5, 3 and 6-month catalogue UCB parameter distributions. Along with a given catalogue, history tree data are produced linking the UCBs of a catalogue to UCBs in the preceding $T_{\rm obs}$ catalogue. The history data will be useful to the observer when determining which catalogue UCBs are potentially confused. Where this is especially useful is for two UCB nearby in frequency and sky location, as discussed in the second corner case of the Appendix.

The overlap integral (\ref{eq:match}) used in the catalogue production step is also used to cross correlate LDC injected waveforms and GBMCMC catalogue waveforms, to determine which catalogue UCBs have a matching LDC injection, with $M_{ij}>0.8$. We see the results of this in the next section where we compare our catalogues to the population of LDC injections to quantify the efficacy of our search.

\begin{table}[h]
\centering
\resizebox{8cm}{!} {
\begin{tabular}{|c|c|}
\hline
Observation time (months)  & GBMCMC catalogue UCBs \\
\hline
1.5 &  1998  \\ \hline
3.0  &  2758  \\ \hline
6.0  &  6196  \\ \hline
12.0 &  10027  \\ 
\hline
\end{tabular}}
\caption{Total number of GBMCMC catalogue UCBs as a function of observation time.}
\label{table1}
\end{table}

\begin{table}[h]
\centering
\resizebox{8cm}{!} {
\begin{tabular}{|c|c|c|c|c|}
\hline
Observation time (months) & S/N$>$7 & S/N$>$17 & S/N$>$40  & S/N$>$100 \\
\hline
\multicolumn{5}{|c|}{$M_{ij}>0.8$} \\ \hline
1.5 & 826 & 181 & 31 & 6 \\ \hline
3.0 & 1819 & 594 & 118 & 11 \\ \hline
6.0 & 4356 & 1481 & 357 & 55  \\ \hline
12.0 & 7255 & 2989 & 780 & 128 \\ \hline
\multicolumn{5}{|c|}{$0.5<M_{ij}<0.8$} \\
\hline
1.5 & 182& 30& 5& 0 \\ \hline
3.0 & 357& 83& 6& 0 \\ \hline
6.0 & 643& 75& 6& 0  \\ \hline
12.0 & 1107 & 115 & 13 & 4 \\ \hline
\end{tabular}}
\caption{Same as \ref{table1} but with S$/$N cuts on $M_{ij}>0.8$ catalogue UCBs (top) and catalogue UCBs with $0.5<M_{ij}<0.8$ (bottom). Catalogue UCB that have $M_{ij}<0.5$ are not included in the S$/$N cut.}
\label{table2}
\end{table}

\begin{table*}[h]
\centering
\resizebox{15cm}{!}{
\begin{tabular}{|c|c|c|c|c|c|c|}
\hline
Observation time & sky localisation  & $\dot f>$0  & $\dot f<$0 & eclipsing & eclipsing with $\dot f>$0 & eclipsing with $\dot f>$0, \\
& 10 sq. deg. &  &  & 70$^\circ<i (\rm{deg.})<$110$^\circ$ & & and well-localised \\
\hline
1.5 & 2 & 279 & 5 & 91 & 16 & 0\\ \hline
3.0  & 3 & 553 & 1 &  372 & 48 & 0\\ \hline
6.0  & 25 & 1950 & 10 & 1027 & 275 & 4\\  \hline
12.0 & 441 & 4239  & 20 & 1873 & 735 & 58 \\ 
\hline
\end{tabular}}
\caption{Same as Tables \ref{table1} and \ref{table2}, but with different cuts on the catalogue UCBs with frequencies above 5 mHz to identify UCBs worthy follow-up analysis by EM observatories. A catalogue UCB is added to each column when at least 90\% of its MCMC samples satisfy the given condition. In the first cut we state the number of UCBs that are `well-localised', with sources contained within a sky location area of 10 deg$^2$. In the next two columns, we have the number of UCB with positive and negative frequency derivatives. The number of eclipsing binaries is shown in column 5, and we further subdivide this category into the last two columns. Eclipsing with GW dominated frequency evolution, and the number of UCBs which are eclipsing and well-localised. From this final category we select a high frequency 12-month catalogue UCB, as an example of a target for follow-up EM observations and archival searches.}
\label{table3}
\end{table*}

\section{\label{sec:ldc_inj}Comparing the GBMCMC catalogues to LDC injections} 
Since we are dealing with a simulated data stream which comes with the parameter values describing the waveforms for every LDC UCB in the data, it is possible to check the efficacy of our search pipeline by cross correlating our catalogue waveforms with the LDC waveforms using Equation (\ref{eq:match}). The methods and results of this process are presented next. These results inform the efficacy of our search pipeline across observing time. Each catalogue UCB is classified as a matched, confused or false alarm detection. The different classifications are explained below.

When the correlation coefficient $M_{ij}$ of a catalogue$-$injection pair exceeds a threshold of 0.8 we regard these as a `matched' pair. There is typically one LDC injection meeting this criterion for the given catalogue UCB. Though it occurs less often, we shall see that it is possible for one catalogue UCB to have a match with two injections. In general, we refer to catalogue UCBs that have a match with one LDC injection, or more, as matched. 

The majority of catalogue UCB that are not matched are classified as confused. Confused catalogue UCB can be further distinguished as two sub-categories of blending: (1) two UCBs each have a positive overlap with the same injection and the sum of the two UCB waveforms has a larger overlap with that injection; (2) a single UCB has a larger overlap with the sum of two injected waveforms than with each injected waveform alone.

\begin{figure}
\begin{center}
\includegraphics[width=3.5in]{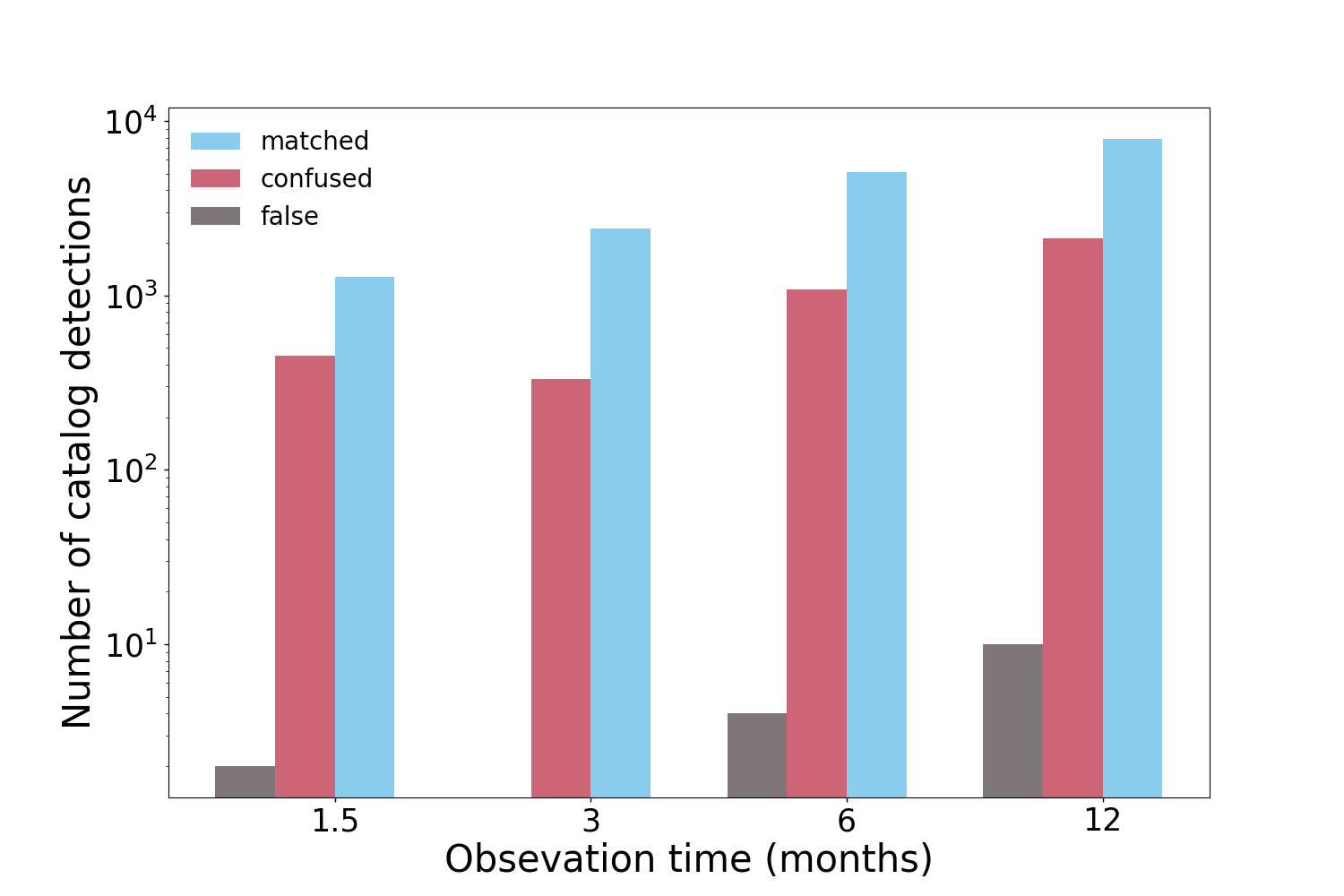}
\end{center}
\vspace*{-5mm}
\caption{Number of catalogue UCBs with observing time, using a logarithmic scale. A catalogue UCB is classified as matched (with a single catalogue UCB and injection satisfying $M_{ij}>0.8$), confused or false.}
\label{histogram}
\end{figure}

\begin{figure}
\begin{center}
\includegraphics[width=3.5in]{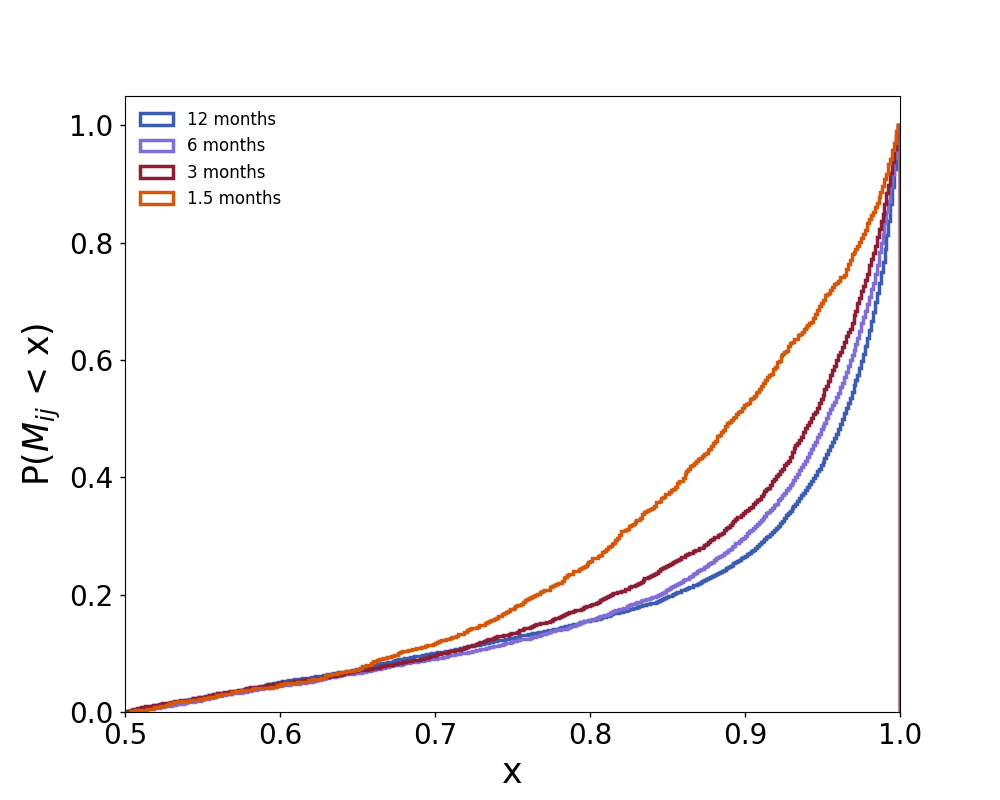}
\end{center}
\vspace*{-5mm}
\caption{Cumulative distribution of catalogue UCB matches over all frequency with correlation values greater than 0.5. Observation times between 1.5 and 12 months are represented.}
\label{cdf}
\end{figure}

A histogram of the number of catalogue UCBs in each of the three categories: matched, confused and false alarm, as a function of observing time is shown in Figure \ref{histogram}. The fraction of catalogue UCBs which have a match with an LDC injection is 0.74, 0.88, 0.82, and 0.79 for the 1.5, 3, 6 and 12-month catalogues, respectively. The fraction of catalogue UCBs which are confused is 0.26, 0.12, 0.18, and 0.21 (for 1.5, 3, 6 and 12-month catalogues). The larger the fraction of confused sources in each catalogue, the smaller the fraction of matched sources. Figure \ref{cdf} shows the cumulative distribution of cross correlation values between the $T_{\rm obs}$-month catalogue UCB and LDC injections. The fraction of catalogue UCBs with match below $M_{ij}=x$ is shown on the vertical axis, and one notices the fraction of 1.5-month catalogue UCBs below the match threshold of 0.8 is significantly larger than the other $T_{\rm obs}$ catalogues. For such a short observation time, a larger fraction of confused sources is expected. 


\begin{figure}[h]
\includegraphics[width=0.5\textwidth]{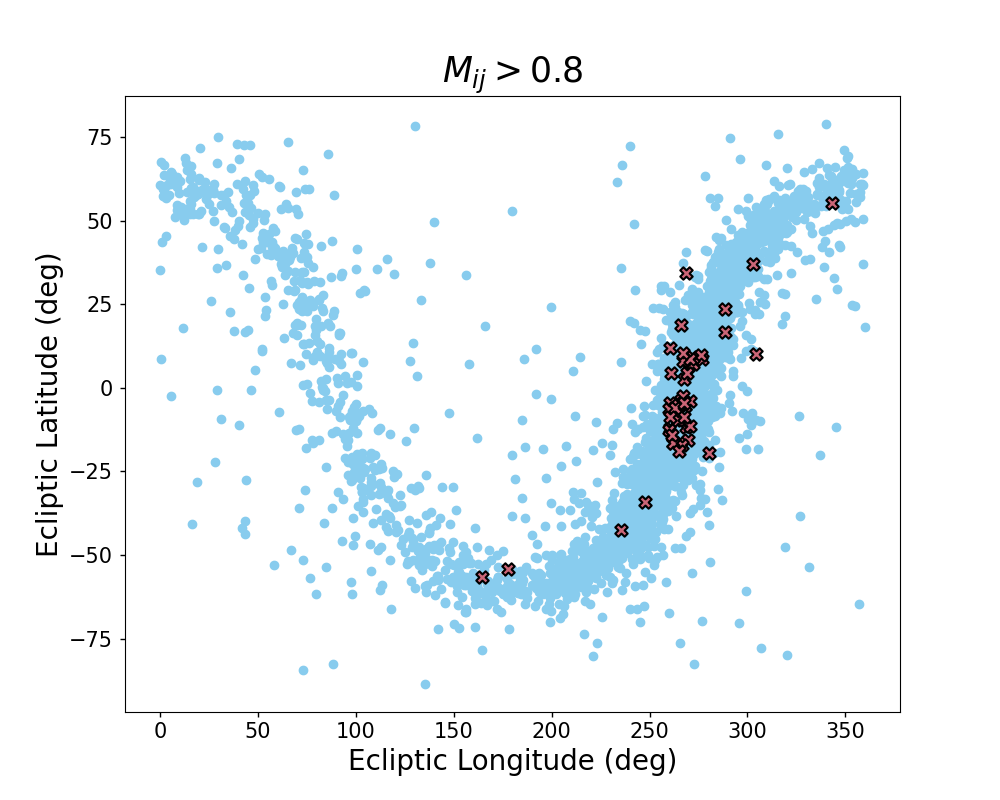} 
\includegraphics[width=0.5\textwidth]{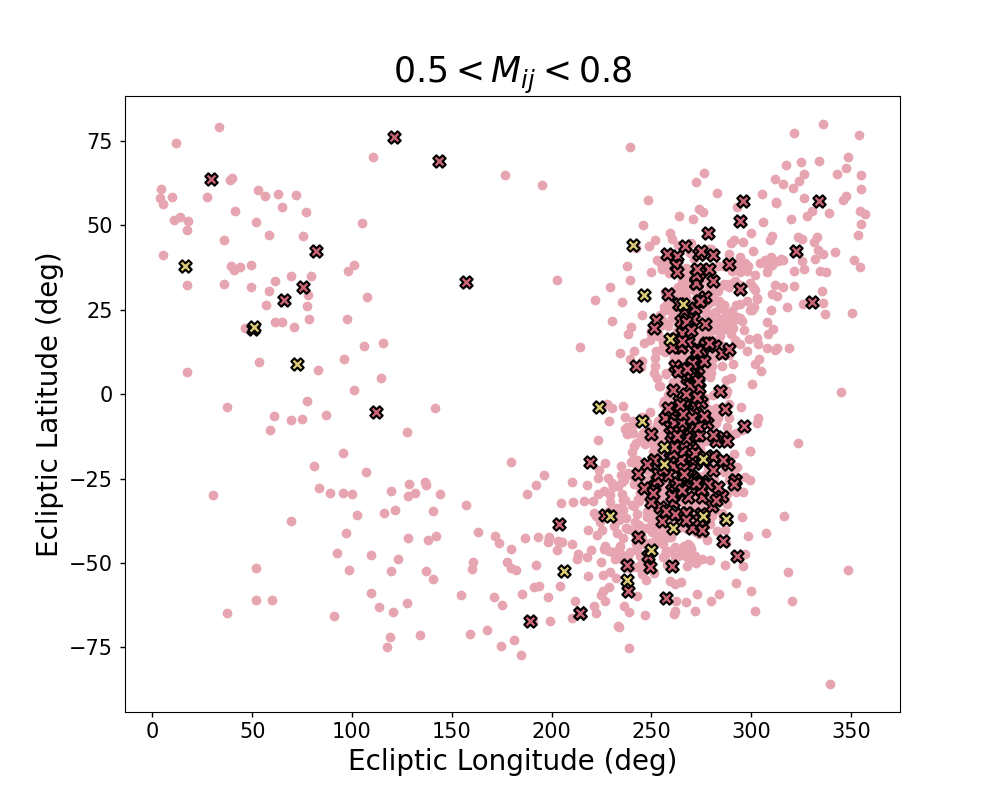} 
\caption{Sky location of blended 12-month catalogue UCBs. In the top and bottom graphs, the sky location of blended 12-month catalogue UCBs are highlighted here as dark pink and yellow crosses with a black border. The underlying distribution of points are all catalogue UCBs meeting a certain correlation threshold. This threshold is $M_{ij}>0.8$ in the top, and the range of $0.5<M_{ij}<0.8$ is used in the bottom graph. The highest concentration of points is near the galactic centre. The overlapping dark pink markers represent catalogue UCBs that have a match with more than one injection (the second type of blending described in Section \ref{sec:ldc_inj}). Each of the, more sparse, yellow markers represent a catalogue UCB that fits the same injection as another catalogue UCB (the first type of blending described in Section \ref{sec:ldc_inj}). For $M_{ij}>0.8$, the first type of blending is absent in the 12-month catalogue. One can see that most of the blended catalogue UCBs are near the galactic centre.}
\label{blended} 
\end{figure}

Even though blending occurs more often for confused catalogue UCB, it is possible for matched UCB to exhibit blending also. For matched 12-month catalogue UCBs we find the second type of blending occurs exclusively. In Figure \ref{blended}, we show the sky location of blended 12-month catalogue UCBs with $M_{ij}>0.8$ and in the bottom graph are the sky locations for blended catalogue UCBs with correlation values in the range $0.5<M_{ij}<0.8$. The blended UCBs are represented as `x'  markers with a black border. In each graph, the underlying distribution of points are the catalogue UCBs meeting a correlation threshold of $M_{ij}>0.8$ and $0.5<M_{ij}<0.8$, for the top and bottom graphs, respectively. 

\begin{figure}[h]
\includegraphics[width=0.5\textwidth]{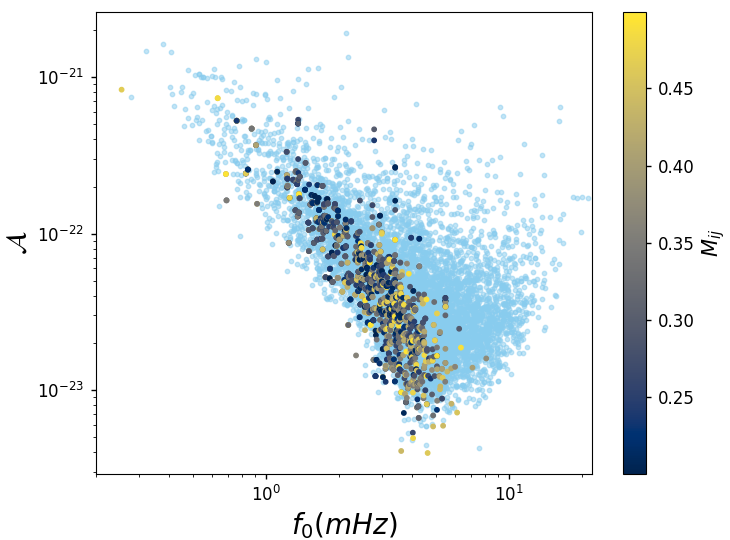}
\includegraphics[width=0.5\textwidth]{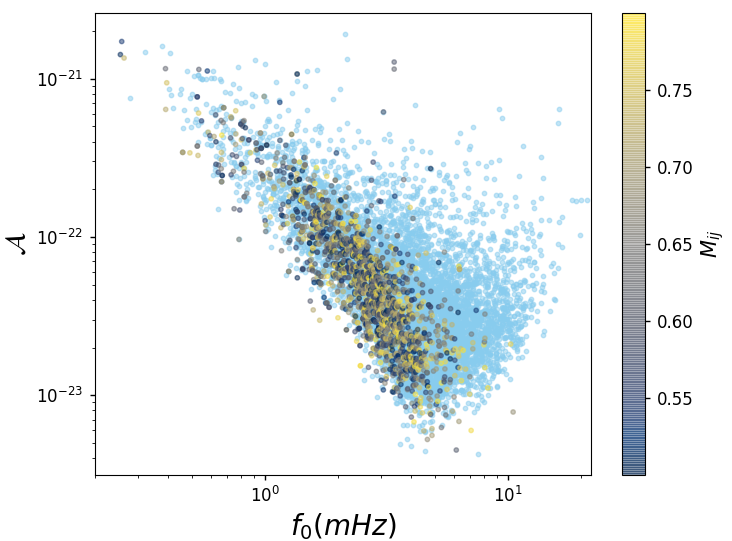}
\caption{Correlation value for confused 12-month catalogue UCBs. In the top graph, the colour-bar shows the correlation value for 12-month catalogue UCBs that are confused with $M_{ij}<0.5$. In the bottom graph, the colour-bar shows the correlation value for 12-month catalogue UCBs that are confused with $0.5<M_{ij}<0.8$. In both graphs, these are plotted over all of the $M_{ij}>0.8$, matching, 12-month catalogue UCBs (light blue).}
\label{no_match} 
\end{figure}

In Figure \ref{no_match} we show the $\amp$-$\f$ plane of all matching catalogue UCBs in blue. Catalogue UCBs that do not have a match are graphed with a colour-bar indicating the largest correlation value with an LDC injection. Each graph displays a different correlation range. The top graph highlights non-matching UCBs with $M_{ij}<0.5$ and the bottom graph highlights non-matching UCBs with $0.5<M_{ij}<0.8$. From both graphs, it is clear that non-matching catalogue UCBs are primarily below 5 mHz. Some of these catalogue UCBs also suffer from blending, that is they also have $M_{ij}<0.5$ with nearby LDC injections that are matched with other catalogue UCBs, which is one reason to set a higher match threshold, in our case to 0.8. 

There is one more type of catalogue UCB that is not classified as confused or matched, according to the categories described above. UCBs are classified as a `false alarm' when no LDC injection exists within 10/$T_{\rm obs}$ of the UCB frequency. However, the false alarms identified in our analysis are due to boundary effects. As $T_{\rm obs}$ becomes larger, the bandwidth of a UCB signal is also wider, leading to long waveform tails. When a long waveform tail also extends beyond the boundaries of an analysis segment, into a neighbouring analysis segment, a false alarm can emerge in the neighbouring segment's catalogue.  In post-processing we explored the highest frequency catalogue UCB false alarm in the 12-month catalogue. This false alarm UCB is located at a $\f \sim 16.64$ mHz. This frequency is on the boundary between two analysis segments and the false alarm UCB waveform template overlaps with the waveform tails of two matched, or recovered, bright UCBs on either side of the boundary. Further examination of the other false alarm UCBs in the catalogue data reveals that each is the symptom of this boundary effect. This symptom is alleviated by allowing communication between neighbouring analysis segments as they run in parallel, such that the residual curve is consistent across the boundary (\cite{Littenberg:2023xpl}). 

\begin{figure*}
\begin{center}
\includegraphics[width=3.5in]{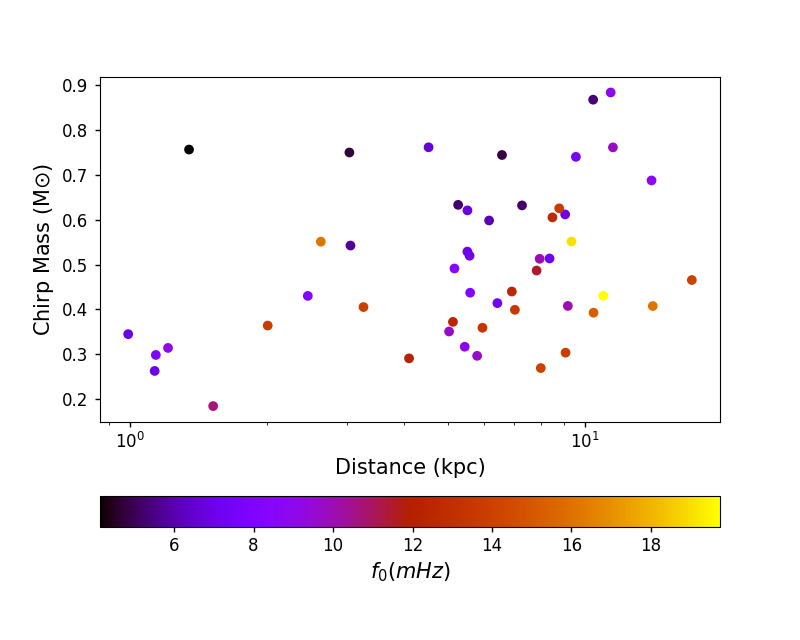}
\includegraphics[width=3.5in]{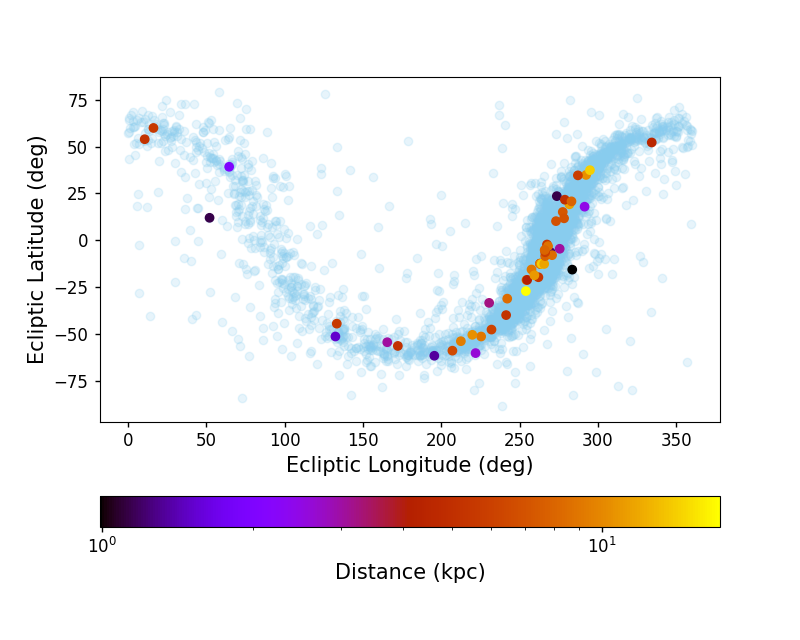}
\end{center}
\vspace*{-5mm}
\caption{Chirp mass versus distance for eclipsing and well-localised catalogue UCB that are matched. The graph on the left shows chirp mass versus distance for all matched 12-month catalogue UCBs that are eclipsing and have positive $\dot f$, coloured by GW frequency. Eclipsing UCB are defined here as having more than 90$\%$ of their inclination angle samples constrained within 70$^\circ<\iota (\rm{deg.})<110^\circ$. The graph on the right shows the ecliptic coordinates of the UCB from the left graph, now coloured by distance. These are plotted over all 12-month catalogue UCB that have $M_{ij}>$0.8 (light blue).}
\label{dist_cuts}
\end{figure*}
\begin{figure*}
\begin{center}
\includegraphics[width=6.5in]{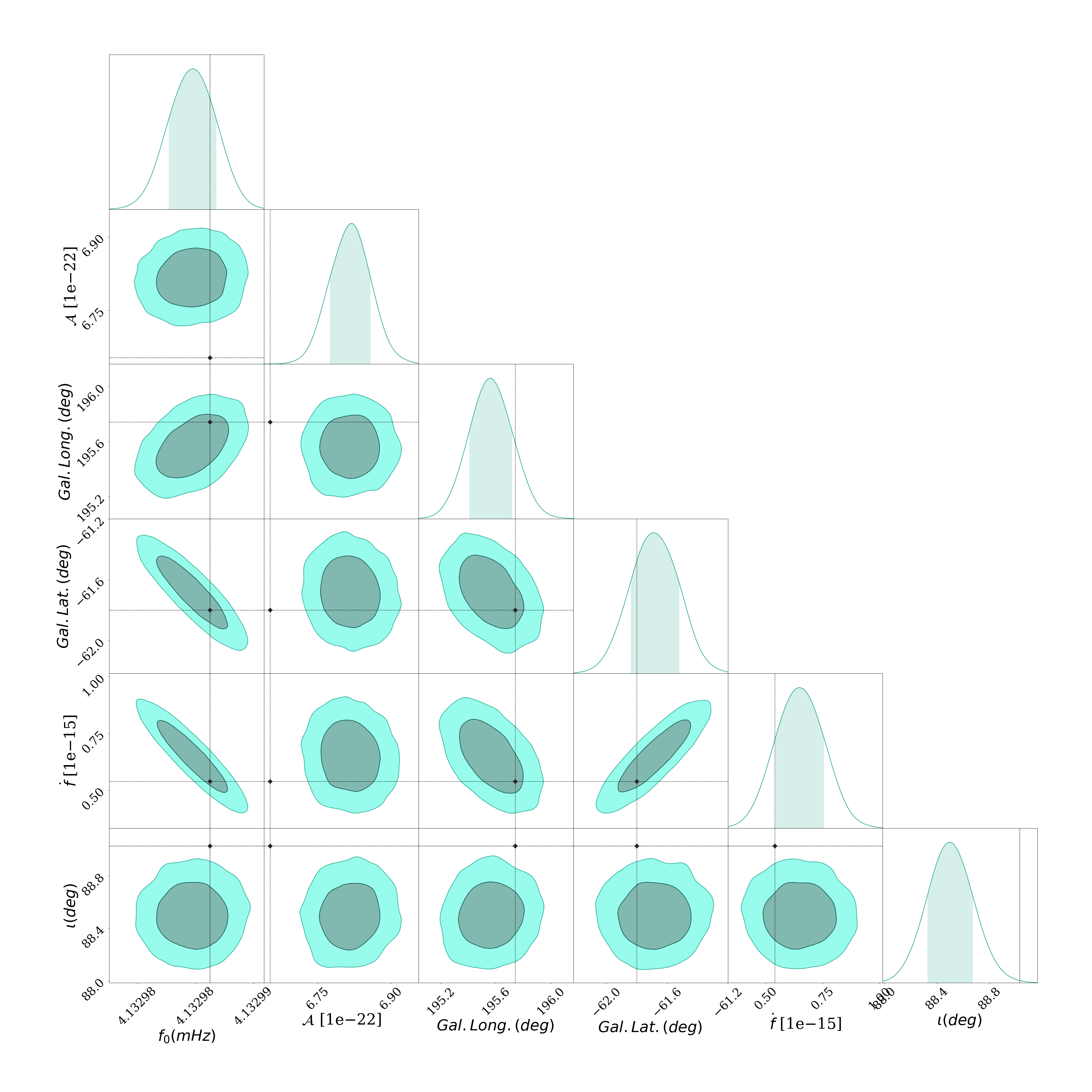}
\end{center}
\vspace*{-5mm}
\caption{The 1 and 2$\sigma$ parameter posteriors for the lowest frequency and 12-month matching catalogue UCB that is eclipsing and localised to within 10 deg$^2$ on the sky. The black diamonds on the 2D posteriors represent the LDC injected parameter values. One can see that the amplitude and inclination angle have not been recovered, and no correlation is visible in the $\amp$-$\iota$ plane. Increasing the number of MCMC steps and searching over a longer observation time are needed to better determine these parameters and recover the expected correlation between them.}
\label{lowest}
\end{figure*}

We identified all 12-month catalogue UCBs that have $M_{ij}>0.8$, are eclipsing and well-localised and graph their chirp mass as a function of distance and sky locations in Figure \ref{dist_cuts}. From the group of matching, eclipsing and well-localised 12-month catalogue UCBs, we selected the lowest frequency source and graphed its parameter posteriors in Figure \ref{lowest}. It has $\iota = 88.50^\circ \pm 0.17^\circ$ and an orbital period of $\sim$8 minutes. The frequency derivative is constrained at 12 months and is positive $\dot f = \left( 63.4^{+11.6}_{-8.8} \right) \times 10^{-17} s^{-2}$. Galactic sky location in degrees is $(\theta, \phi) = (-61.70^{+0.18}_{-0.13}, 195.55^{+0.16}_{-0.13})$. The distance to the system is relatively close at $1.37^{+0.25}_{-0.22}$ kpc, and its chirp mass is $0.751^{+0.074}_{-0.078}\ M_{\odot}$. Even though this recovered UCB has a high correlation value ($M_{ij} = 0.9995$), the joint $\amp$-$\iota$ posterior to the injected amplitude and inclination reveals that these parameters were not accurately recovered. No correlation is visible in the $\amp$-$\iota$ plane, indicating insufficient sampling of the parameter space. Moreover, this 12-month UCB system is also found in the 1.5, 3 and 6-month catalogues with correlation values $>$ 0.8. For reference, this low frequency catalogue UCB is overlapping with more than $\sim$ 30 UCBs. All of these are packed within a frequency range of only $\pm 2^8/T_{\rm 12mo}$. There are 12-month catalogue UCB with even lower frequencies, and with $S/N\gtrsim 8$, but have poorly constrained sky location and distances due to confusion with tens of thousands of unresolved UCB. More observation time is required to determine if the lowest frequency systems are suitable for EM follow-up observations. 

\begin{figure*}[]
\begin{center}
\includegraphics[width=6.5in]{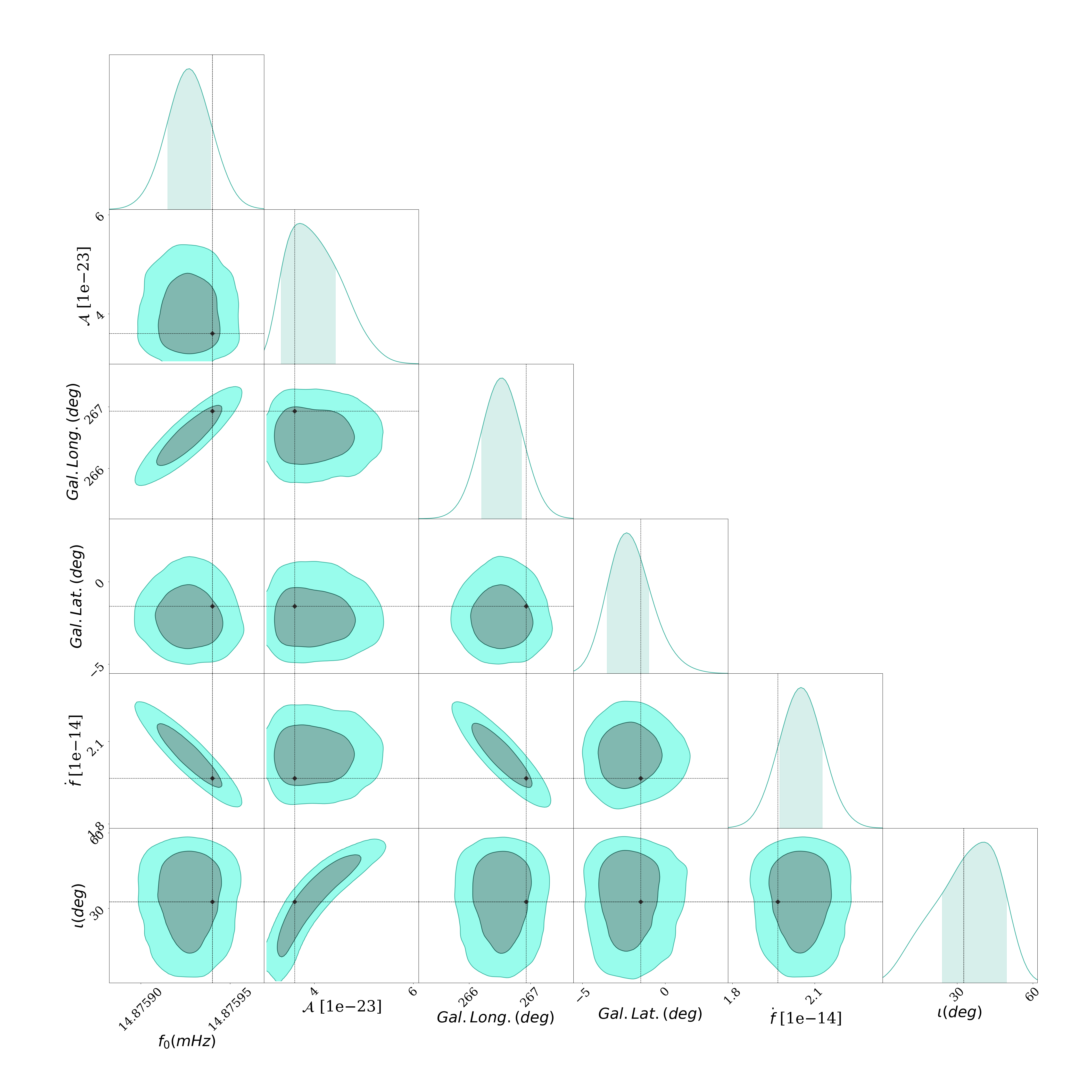}
\end{center}
\vspace*{-5mm}
\caption{The 1 and 2$\sigma$ parameter posteriors for a distant ($D_L=16.1\pm 2.4$ kpc) and 12-month matching catalogue UCB that is localised to within 22 deg$^2$ on the sky. The black diamonds on the 2D posteriors represent the LDC injected parameter values. In galactic coordinates, the source line of sight is $\sim 3$ degrees offset from the galactic centre. At a distance of $\sim 16$ kpc, this places the UCB behind the galactic centre in a region inaccessible to optical telescopes.}
\label{distant1}
\end{figure*}

Posteriors for a distant ($D_L=16.1\pm 2.4$ kpc) and reasonably well-localised (22 deg$^2$) UCB in the 12-month catalogue, with a correlation value M=0.99 and S/N of 36, are shown in Figure \ref{distant1}. One sees that GW parameters $\amp$ and $\iota$ are strongly correlated. This high S/N 12-month UCB system is matched in the 1.5, 3 and 6-month catalogues, with S/N = 2, 18, 22, respectively. The 3D location of this binary system places it in the part of the Milky Way that is inaccessible to optical telescopes due to intervening dust and gas obscuration, called the Zone of Avoidance. LISA will complement radio and infrared surveys in providing a new view to this part of the Milky Way Galaxy. 

In the Appendix, we put the data obtained from comparing the GBMCMC catalogues to the injections to further use by examining a few 12-month frequency segments that contain confused catalogue UCBs. In particular, we explore catalogue UCBs that have a match at six months but do not appear as matched UCBs in the 12-month catalogue. Each corner case involves blended 12-month catalogue UCBs of the first type discussed in the beginning of this section.

\section{\label{sec:discussion}Discussion}

This paper is a report on the analysis of the LISA Data Challenge $Radler$ data with the GBMCMC code. This work is a necessary step in our efforts to prepare for the global fit required for the LISA mission. The LDC simulated data stream contains millions of white dwarf binary signals. We divided the time series data into different observation time increments, $T_{\rm obs}$, of 1.5, 3, 6 and 12 months and produce catalogues for each $T_{\rm obs}$ by performing a global fit to the resolvable binaries. 

GBMCMC, is a trans-dimensional reversible jump MCMC algorithm with parallel tempering. The MCMC sampling was done in frequency, where the full frequency band has been divided into a total of 3317 frequency segments. Bayesian inference is used to select the highest evidence model to build catalogues for each of the 3317 frequency segments as a function of observing time. These are all combined for each observing time to create 1.5, 3, 6 and 12-month catalogues.

The UCB catalogue waveforms are then cross correlated with the known LDC injected waveforms to determine the efficacy of our search pipeline. For each observation time, we quantify the number of matching, confused, and false detections in the catalogues. We recover more than 10,000 binaries after 12 months of observing. We found that 7,255 of these 12-month catalogue UCB have a match with an LDC injection (with a correlation value $M_{ij}>$ 0.8) and have S/N $>$ 7. Of these, there are 128 UCBs with a S/N $>$ 100.

We identify two interesting corner cases for in-depth follow-up analysis. For the first Appendix corner case we investigate two confused 12-month catalogue UCBs, the sum of which are a match with a single LDC injection. Moreover, the two catalogue UCBs have a common ancestor in the 6-month catalogue. Namely, they both have a match value greater than 0.5 with a single 6-month catalogue UCB. This indicates that the 12-month MCMC analysis has not converged. Smaller time-jumps between catalogues is a target for future investigation. 

In the second corner case, we examined a different type of confused UCB in the 12-month catalogue. There are two UCBs occupying the same region of the $\amp$-$\f$ and $\theta$-$\phi$ planes. In this region of parameter space, there are two LCD injections. After a 12-month analysis neither of the injections are separately a match with the confused UCBs; however, the sum of the UCB waveforms is a match with the sum of the LDC injected waveforms. Similar to the previous corner case, we find the 12-month MCMC analysis for this analysis segment has not converged. Additional incremental analysis is necessary between the 6-month and 12-month catalogues to disentangle these two LDC injections. 

These corner cases are part and parcel of a broader discussion of a strategy for creating and publishing catalogues. Namely, how often will the catalogues be updated and released, and the type and format of information dispensed in the catalogues. Alerts and low-latency analysis and outputs will be informed by our time-evolving catalogues. Further in-depth analysis of blended catalogue UCBs in data challenges will be essential for answering some of these questions. Checking the convergence of analysis segments that have catalogue UCBs which share the same parent or are close in sky location and frequency-amplitude will be necessary. Accurate UCB science and low frequency GW science in general depend heavily on validation of the data products output from search pipelines.

We make one last comment regarding future work. Binary UCBs that make up the $Radler$ LDC simulated data stream have zero eccentricity, which is likely the case for most resolvable binaries. However, eccentric UCBs do exist, of course, and  one of the future upgrades to GBMCMC will be to incorporate eccentric UCB models into GBMCMC. The harmonics of low frequency eccentric binaries will be put to use in the search routine. Holding an arbitrary number of parameters fixed at values determined, for example, with radio, optical, X-ray or $\gamma$-ray observations of known binaries (\cite{Littenberg_2019b}) will be of particular interest when checking to see if a newly timed pulsar is observable in the LISA band. 

\bibliographystyle{aa}
\bibliography{refs}

\begin{appendix}
\section{\label{sec:app}Analysis of corner cases}

The following `corner-case' investigations are an exploration of confused regions of parameter space. The cases treated here are limited to catalogue UCBs with frequencies above 5 mHz. The higher frequency segments are of particular interest for post analysis, as we expect the confusion noise to be lower compared to the lower frequency segments. Anomalies in this part of the spectrum will help in pipeline development. 

Table \ref{Table4} is an overview of the first two cases, where the central frequency of the analysis segment, the number of injections, catalogue UCBs and matched UCBs are shown. Three, 6 and 12 month results are displayed for Case 1, and a 24 month analysis is included for Case 2. 

\subsection{Case I}
For the first case, our focus is a catalogue UCB that has a match ($M_{ij}>0.8$) with an LDC injection (INJ005892423) at 6 months. The frequency segment containing this injection was identified as a useful corner case, because the injection was not recovered in the 12-month catalogue. Furthermore, two blended catalogue UCBs (LDC0058924130 and LDC0058923825) appear in the 12-month catalogue within 10/$T_{\rm obs}$ of the injection frequency, indicating a problem with convergence. 
 
We quantify the degree of blending for the two confused 12-month catalogue UCBs. Using Equation (\ref{eq:match}), catalogue UCBs LDC0058924130 and LDC0058923825 respectively have correlation values of 0.73 and 0.67 with LDC injection INJ005892423. Moreover, from the history data, we find that the two 12-month catalogue UCBs share the same parent in the 6-month catalogue. This is also a clue that these two UCB templates are fitting the same feature in the data. The sum of the two templates waveforms has a correlation value of 0.9 with INJ005892423. 

This first case is an example of poor convergence of the sampler. Two templates fitting one injected signal is not a parsimonious solution; on the other hand, one template fitting two or more injections can be the correct answer in a Bayesian sense. 

The next avenue of investigation of sampler convergence would be to adjust the cadence of catalogue creation. The jump between the 6 and 12 month analysis is potentially too large for the `posterior to proposal' method (\cite{lit20}). We discuss this issue in Section~\ref{sec:discussion}.

In the next corner case, we find that for confused regions of frequency-amplitude and sky location parameter space issues of convergence also emerge.

\begin{figure*}
\begin{center}
\includegraphics[width=0.45\textwidth]{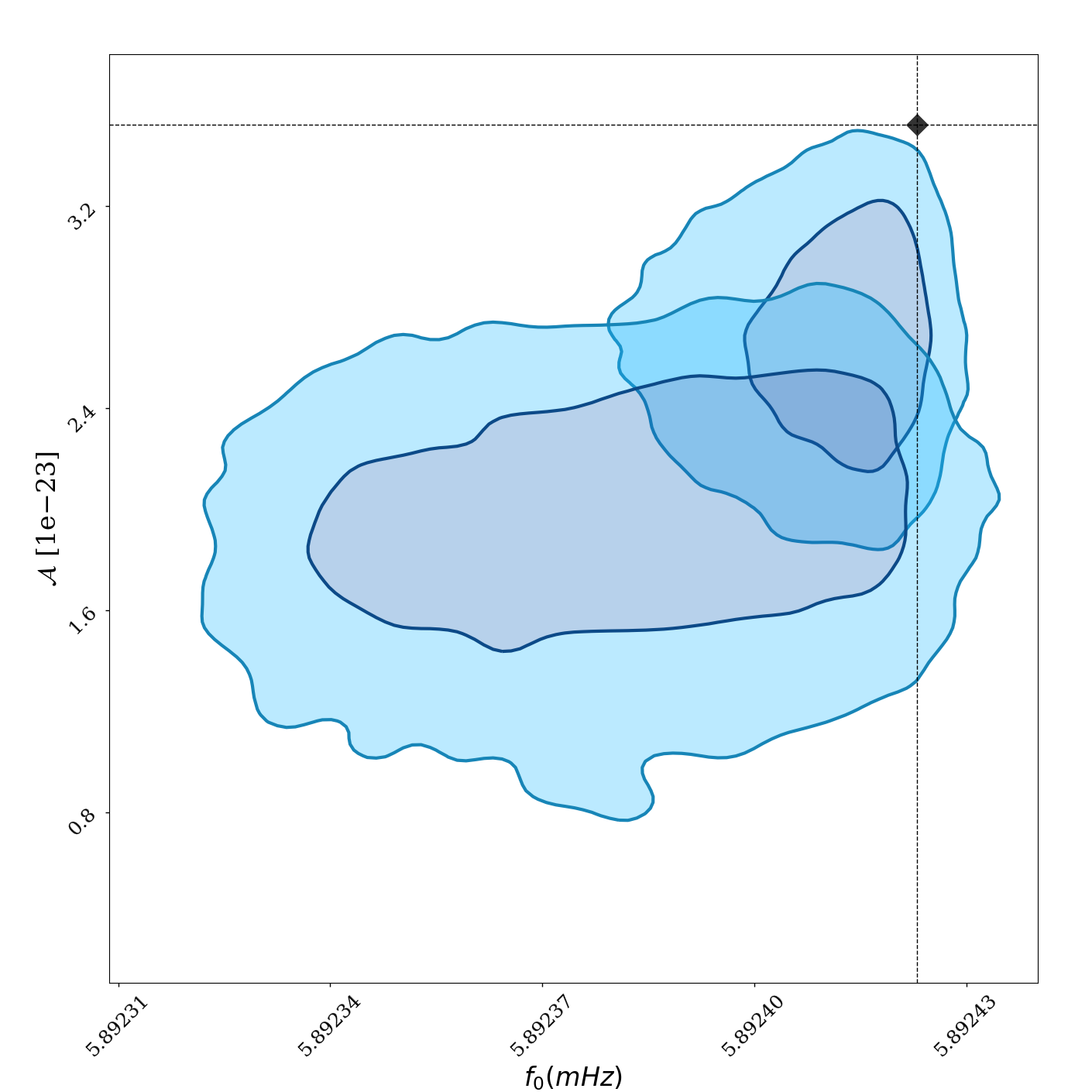} 
\includegraphics[width=0.45\textwidth]{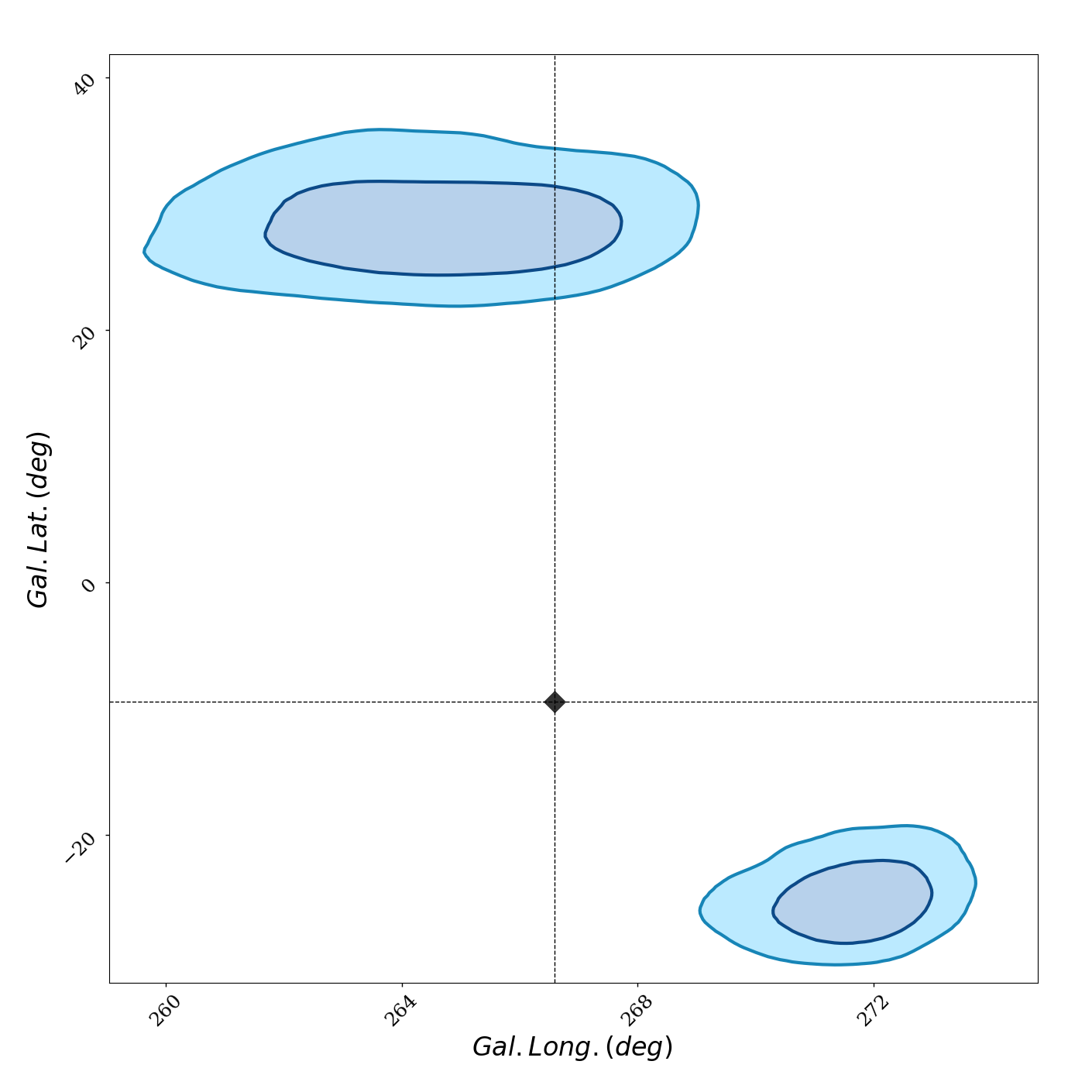} 
\end{center}
\vspace*{-5mm}
\caption{Corner case I: The 1 and 2$\sigma$ posteriors for the $\f$-$\amp$ and sky location planes for two non-matching catalogue UCBs, for the 24 month analysis. The injected parameter values are displayed as black markers. }
\label{history1}
\end{figure*}

\begin{figure*}
\begin{center}
\includegraphics[width=0.6\textwidth]{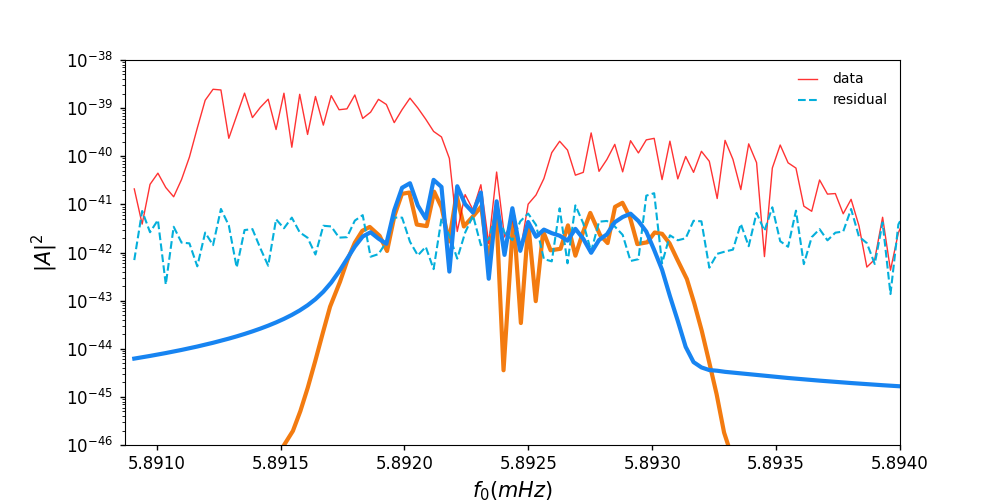} 
\end{center}
\vspace*{-5mm}
\caption{The PSD for corner case 1. The A-channel power spectral density of the sum of confused catalogue UCB waveforms (LDC0058924130+LDC0058923825) is in orange, and the LDC injected waveform (INJ005892423) that matches with the summed catalogue waveforms is the solid blue curve. The original data are in red, and residual is plot as a dashed blue curve. One can see from the red curve that there are at least two high S/N UCBs on either side of the graphed waveforms. Except for INJ005892423, all injected waveforms have been recovered.}
\label{history2}
\end{figure*}

 \begin{table*}
\begin{center}
\resizebox{15cm}{!}{
\begin{tabular}{|c|c|c|c|c|}
  \hline
Case &  Central frequency  & Number of LDC & Number of catalogue & Number of matches   \\
& of segment (mHz) &  injections    & UCBs     &        \\
\hline
& & & \multicolumn{1}{|c|}{\ \hfill 3mo\hfill\  \vline\hfill \ 6mo\hfill\  \vline\hfill \ 1yr\hfill\ \vline\hfill \ 2yr\hfill\ \null} & \multicolumn{1}{|c|}{\ \hfill 3mo\hfill\  \vline\hfill \ 6mo\hfill\  \vline\hfill \ 1yr\hfill\ \vline\hfill \ 2yr\hfill\ \null}               \\
\hline
 1. & 5.89600 & 7 & \multicolumn{1}{|c|}{\ \hfill 2\hfill\  \hfill \ 6\hfill\  \hfill \ 7\hfill\ \hfill \ --\hfill\ \null} & \multicolumn{1}{|c|}{\ \hfill 2\hfill\  \hfill \ 6\hfill\  \hfill \ 5\hfill\ \hfill \ --\hfill\ \null}                                       \\
  \hline
 2. & 5.21647 & 13 & \multicolumn{1}{|c|}{\ \hfill 4\hfill\  \hfill \ 11\hfill\  \hfill \ 12\hfill\ \hfill \ 12\hfill\ \null} & \multicolumn{1}{|c|}{\ \hfill 3\hfill\  \hfill \ 11\hfill\  \hfill \ 10\hfill\ \hfill \ 10\hfill\ \null}                            \\
  \hline
\end{tabular}}
\caption{Overview of Appendix corner cases, corresponding to Figures \ref{history1}, \ref{history2}, \ref{history3} and \ref{history4}.}
\label{Table4}
\end{center}
\end{table*}

\subsection{Case II}
\begin{figure*}
\begin{center}
\includegraphics[width=0.45\textwidth]{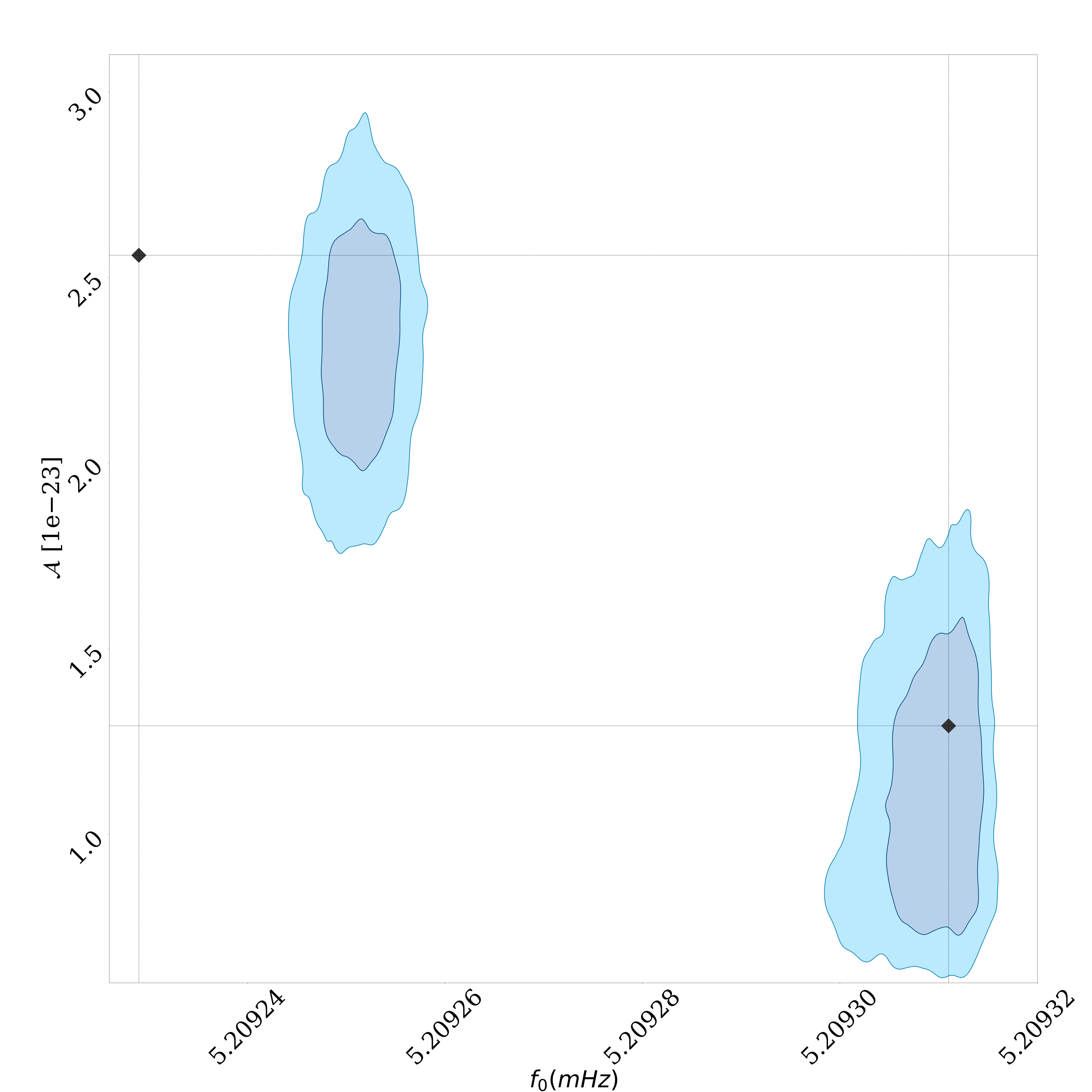} 
\includegraphics[width=0.45\textwidth]{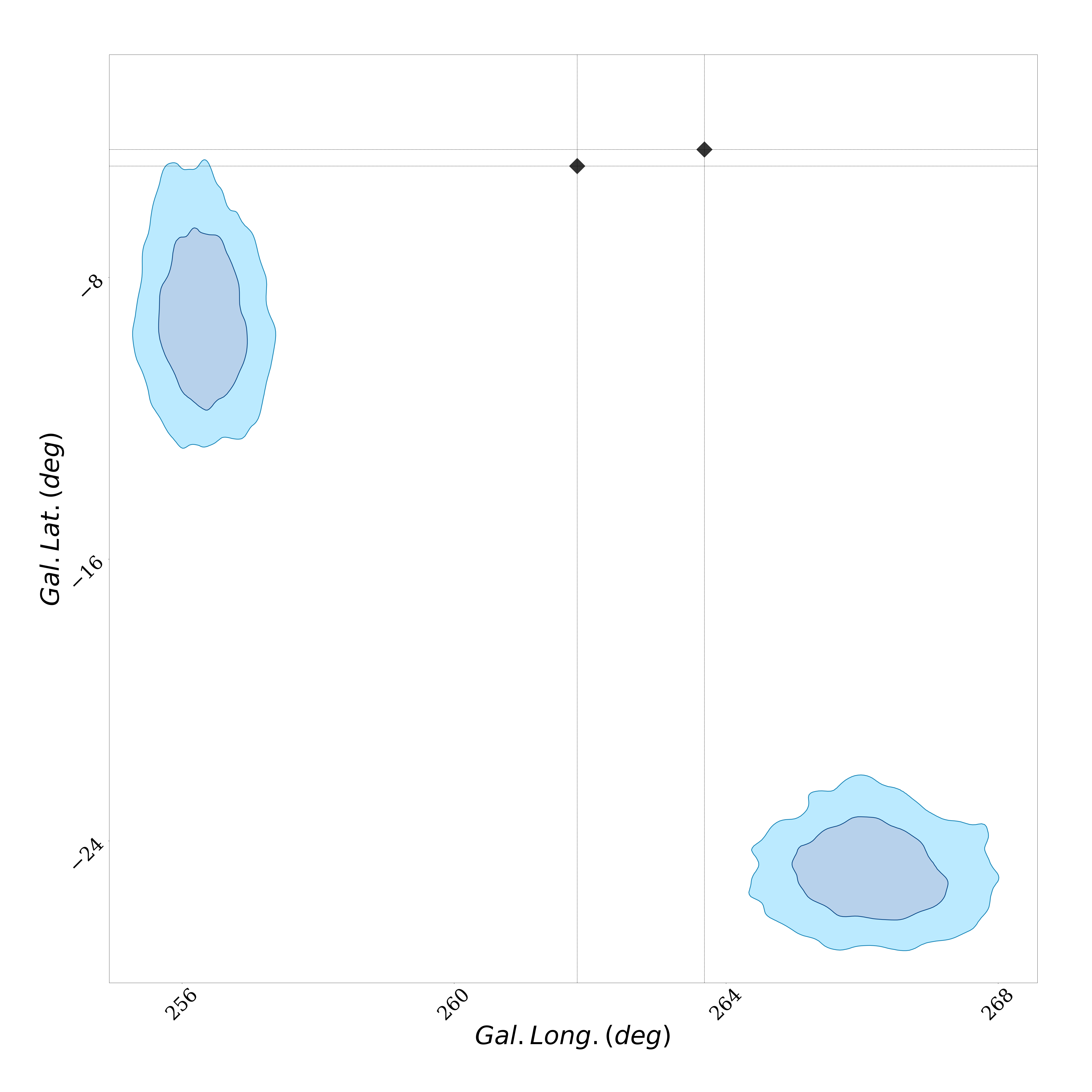}
\end{center}
\vspace*{-5mm}
\caption{Corner case 2: The 1 and 2$\sigma$ posteriors for the $\f$-$\amp$ and sky location for two non-matching catalogue UCBs, for the 24 month analysis. The injected signal values are the black markers.}
\label{history3}
\end{figure*}

\begin{figure*}
\begin{center}
\includegraphics[width=0.6\textwidth]{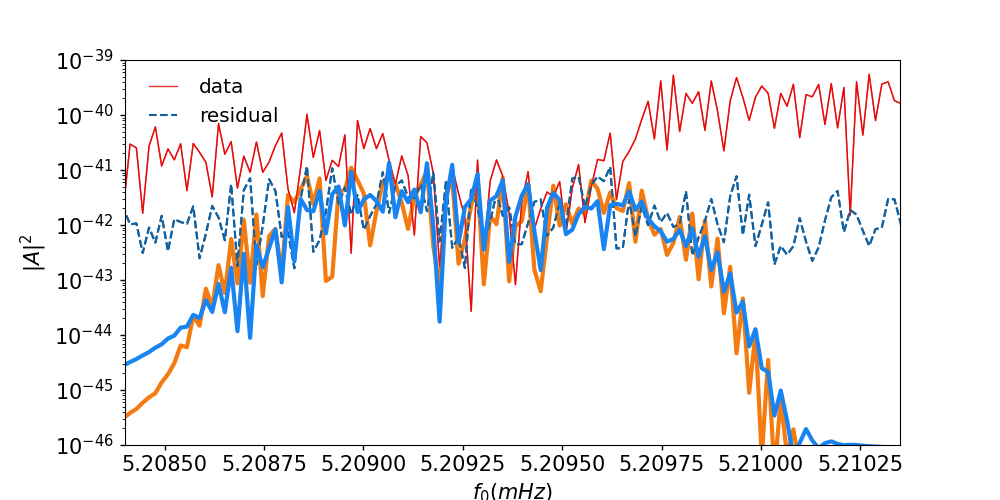} 
\end{center}
\vspace*{-5mm}
\caption{The PSD for corner case 2. The solid blue curve is the sum of LDC waveforms INJ00520923 and INJ00520931. The waveform sum of the two catalogue UCBs in orange is a match with the sum of LDC waveforms. No other combination is a match. One can see from the red curve that there are high S/N UCBs on either side of the graphed waveforms. With the exception of INJ00520923 and INJ00520931, all injected waveforms have been recovered.}
\label{history4}
\end{figure*}

Similar to corner case I, we focus on a catalogue detection that has a match with an LDC injection at 6 months but does not have a corresponding match within the 12-month catalogue. Moreover, this case also involves two 12-month catalogue UCBs which are blended. However, in this analysis segment there are two nearby LDC injections in the $\amp$-$\f$ plane. These two injections are nearby in sky location also. The sum of two 12-month catalogue UCBs is a match with the sum of the two LDC injections, with a correlation value of 0.83, with no other combination of waveforms having $M_{ij}>0.8$. We refer to the two injections as INJ00520923 with 24-month S$/$N = 25 and INJ00520931 with 24-month S$/$N = 23. For the same frequency segment, we performed an analysis of 24 months of data and found the two injections are again not matched. As in the 12 month analysis, we find two UCBs that are attempting to fit the two injections. The sum of the two 24-month catalogue UCB waveforms is a match with the sum of these two injected waveforms, again with a match of $0.83$ and no other matching combination. See Figures \ref{history3} and \ref{history4}.

In an effort to explore GBMCMC convergence, we re-analysed this 24 month frequency segment nine additional times with 100,000 MCMC steps and 10 additional times with 200,000 steps. Another run with 300,000 steps was also completed. 
A summary of the MCMC results is as follows. For 100,000 MCMC steps, we find that that the number of catalogue UCBs from each run is inconsistent. As suspected, the GBMCMC sampler has not converged. Doubling or tripling the number of steps does not lead to convergence either.  

Unlike corner case I, the blended 12-month catalogue UCBs do not share the same parent in the 6-month catalogue. However, they are nearby in sky location, as well as frequency, a potential clue that these two catalogue UCBs may be confused and require additional scrutiny. As in the previous case, adjusting the cadence of catalogue creation is a possible solution.

\end{appendix}

\section*{Acknowledgments}
KL's research was supported by an appointment to the NASA Postdoctoral Program at the NASA Marshall Space Flight Center, administered by Universities Space Research Association under contract with NASA. KL thanks Robert Main for helpful comments.

\end{document}